\documentclass[10pt,twocolumn,showpacs,showkeys,preprintnumbers,amssymb,aps,superscriptaddress,prb]{revtex4-2}

\usepackage{color}
\usepackage{amsmath}
\usepackage{graphicx}
\usepackage{hyperref}

\usepackage{orcidlink}

\begin{document}
\title{Quantum Entanglement Generation in the Heterometallic Ni$^\text{2+}_4$Gd$_4^\text{3+}$ Complexes }
 
\author{Hamid Arian Zad\orcidlink{0000-0002-1348-1777}}
 \email{Corresponding author: hamid.arian.zad@upjs.sk}
 \address{Department of Theoretical Physics and Astrophysics, Faculty of Science, P. J. \v{S}af{\'a}rik University, Park Angelinum 9, 040 01 Ko\v{s}ice, Slovak Republic}

\author{Michal Ja{\v s}{\v c}ur\orcidlink{0000-0003-0826-1961}}
\address{Department of Theoretical Physics and Astrophysics, Faculty of Science, P. J. \v{S}af{\'a}rik University, Park Angelinum 9, 040 01 Ko\v{s}ice, Slovak Republic}

\author{Azam Zoshki \orcidlink{0000-0002-0023-6094}}
\address{Department of Theoretical Physics and Astrophysics, Faculty of Science, P. J. \v{S}af{\'a}rik University, Park Angelinum 9, 040 01 Ko\v{s}ice, Slovak Republic}

\author{Ralph Kenna \orcidlink{0000-0001-9990-4277}}
\address{Centre for Fluid and Complex Systems, Coventry University, Coventry, CV1 5FB, United Kingdom}
\address{$\mathbb{L}4$ Collaboration \& Doctoral College for the Statistical Physics of Complex Systems, Leipzig–Lorraine–Lviv–Coventry, Europe
	}

\author{Nerses Ananikian\orcidlink{0000-0001-8974-1275}}
\address{
	A.I. Alikhanyan National Science Laboratory, 0036, Yerevan, Armenia}



\date{\today}

\begin{abstract}
We investigate various types of quantum entanglement in the octanuclear heterometallic \(3d/4f\) complexes denoted as Ni$^{2+}_4$Gd$^{3+}_4$ under an external magnetic field, using the { exact diagonalization approach}. These molecular magnets, which can be effectively described by Heisenberg spin models, consist of two identical \{Ni$^{2+}_2$Gd$^{3+}_2$\} cubane subunits bridged by acetate and hydroxide ligands. Our analysis reveals that their magnetization exhibits intermediate plateaus at low temperatures, indicating distinct ground states characteristic of Ni-containing compounds.
Using negativity as a measure of quantum entanglement, we examine the influence of single-ion anisotropy and magnetic field on tetrapartite, bipartite, 1--3 tangle, and 2--2 tangle entanglements in two families of Ni$^{2+}_4$Gd$^{3+}_4$ complexes: \textbf{(1)} without anisotropy and \textbf{(2)} with anisotropy. Complex~\textbf{(1)} exhibits strong bipartite entanglement between Ni ions, which persists up to \(T \approx 3.0\,\text{K}\) and \(B \approx 4.0\,\text{T}\), but shows significantly weaker tetrapartite entanglement and vanishing bipartite entanglement between Gd$\cdots$Gd and Ni$\cdots$Gd pairs. In contrast, complex~\textbf{(2)} displays nonzero and sizable values for all types of entanglement considered.
These findings emphasis the crucial role of single-ion anisotropy in generating and shaping the entanglement landscape of heterometallic Ni$^{2+}_4$Gd$^{3+}_4$ complexes. Notably, we find that the 1--3 tangle entanglement between a Ni ion and the remaining sites in a cubane unit serves as a reliable indicator of ground-state phase transitions, exhibiting distinct changes across phase boundaries irrespective of the presence of single-ion anisotropy.

\end{abstract}

\pacs{Valid PACS appear here}
\maketitle


\section{Introduction} \label{sec:introduction}

Molecular magnets with interacting spins \cite{Kahn, Chiesa2024, Chilton2013, Law2011, Miller2002} provide a powerful platform for advancing quantum science and technology. Their complex low-energy spectra make them promising candidates for quantum information storage and processing \cite{Gatteschi2006, Carretta2021}. This unique property enables their application as qudits, extending the capabilities of quantum logic beyond conventional qubit-based architectures.  
One particularly intriguing aspect of low-dimensional Heisenberg spin systems is the emergence of commensurate magnetization plateaus related to the distinct quantum states that arise purely from quantum mechanical effects. These plateaus become especially pronounced at low temperature, where the suppression of thermal fluctuations allows quantum states to manifest clearly \cite{Eggert1994, Eggert1996, Motoyama1996, Oshikawa1999, Affleck1997, Kikuchi2005, Hida1994, Oshikawa1997, Leiner2018, ArianPhysA, ArianZad2025, Balents2010, Balents2017}.  

The renewed interest in mixed $3d/4f$ metal cluster chemistry is driven by the discovery of molecular nanomagnets and their extraordinary magnetic properties which arise from the interplay between transition metal ($3d$) and lanthanide ($4f$) ions. Unlike homometallic $3d$ clusters, the design of $3d/4f$ molecular clusters requires the strategic incorporation of lanthanide ions, with Gd$^\text{3+}$ being a promising candidate due to its highly isotropic nature and large spin moment \cite{Zhao2014,Mukhopadhyay2004,Khuntia2011,Xie2013,BiswasNJC2020}. The significant exchange interactions between $3d$ and $4f$ metal centers lead to tunable magnetic behaviors, making these systems ideal for both fundamental research and practical applications.  
Among these materials, heterometallic  Ni$^\text{2+}_4$Gd$_4^\text{3+}$ complexes \cite{Biswas2020,Kalita2018, Pasatoiu2014,Biswas2016} have drawn considerable attention due to their intriguing magnetic properties. 
AC magnetic susceptibility measurements revealed fast relaxation dynamics in these complexes, attributed to quantum tunneling of magnetization (QTM). As a result, neither of these compounds exhibits the slow relaxation characteristic of single-molecule magnets (SMMs). The absence of magnetic hysteresis makes these non-SMM systems particularly valuable for quantum statistical physics, as they can be effectively analyzed using statistical methods.  
The precise engineering and control of magnetic interactions at the molecular level make these clusters promising candidates for advanced molecular spin-based devices. Their potential applications can span cutting-edge technologies, including quantum information processing \cite{Troiania2011}, spintronics \cite{Wernsdorfer2008}, and the development of quantum states for next-generation quantum computing \cite{Bagai2009, Christou2000, Hill2003}.
 
Low-spin molecular nanomagnets, particularly those exhibiting antiferromagnetic interactions or anisotropy, provide an optimal setting for studying quantum entanglement \cite{Wootters1996,Wootters1997,Wootters1998,Aldoshin2014,Peres1996,Vidal2002,Horodecki2009}. Quantum entanglement is a fundamental property required for utilizing molecular nanomagnets as spin cluster quantum states \cite{Chiesa2024,Troiania2011,Aldoshin2014}. Negativity is a widely used measure of quantum entanglement \cite{Peres1996,Vidal2002,Horodecki2009}, applicable to mixed states of bipartite and multipartite systems, such as qutrit-qutrit and qutrit-qutrit-qudit-qudit configurations.  
The study of magnetic properties of the polycrystalline samples Ni$^\text{2+}_2$Gd$_2^\text{3+}$ such as {\bf (1)}
$[\text{Ni}_4\text{Gd}_4(\text{HL})_4(\mu_2-\text{OH})_2(\mu_3-\text{OH})_4(\mu-\text{OOCCH}_3)_8] \cdot (\text{NO}_3) \cdot 12\text{H}_2\text{O}$  \cite{Biswas2020}  
indicated a ferromagnetic ${\mathrm{Gd}\cdots\mathrm{Gd}}$ and an antiferromagnetic ${\mathrm{Ni}\cdots\mathrm{Ni}}$ interaction.  Similarly, a related study on {\bf (2)} 
$[\text{Ni}_4\text{Gd}_4(\mu_2-\text{OH})_2(\mu_3-\text{OH})_4(\mu-\text{OOCCH}_3)_8(\text{LH}_2)_4]$ \cite{Kalita2018}  
revealed a ferromagnetic interaction between the lanthanide and nickel centers, with an estimated exchange coupling of $J_{\mathrm{Ni}\cdots\mathrm{Gd}} = +0.86\;\text{cm}^{-1}$. However, these complex does not exhibit SMM behavior due to weak $\text{Ni} \cdots \text{Gd}$ and $\text{Gd} \cdots \text{Gd}$ interactions.

In this work, we study a Heisenberg spin model representing the $\text{Ni}_4^{2+}\text{Gd}_4^{3+}$ complexes and perform a rigorous analysis of their low-temperature magnetization, as well as various quantum entanglement measures, including tetrapartite and bipartite negativities, and 1--3 and 2--2 tangles that characterize the entanglement among all or selected subsets of spins within each cubane unit.
These complexes consist of two identical butterfly-shaped heterometallic $\text{Ni}_2^\text{2+}\text{Gd}_2^\text{3+}\text{O}_4$ distorted cubanes, weakly connected via acetate and hydroxide bridging ligands. The presence of non-negligible single-ion anisotropy in Ni ions introduces additional complexity, as this property competes with exchange interactions in determining the quantum characteristics of these molecular magnets. 
For each distorted cubane unit, \(\text{Ni}_2^{2+}\text{Gd}_2^{3+}\), the Hilbert space dimension is \((2S_\mathrm{Ni}+1)^2 \times (2S_\mathrm{Gd}+1)^2 = 3^2 \times 8^2=576\), corresponding to two spin-1 (Ni) and two spin-7/2 (Gd) ions. While the Hamiltonian cannot be solved analytically, its moderate size allows us to perform full numerical diagonalization without approximation. With the fact that the inter-cubane interactions in these complexes are significantly weaker than the intra-cubane exchange interactions, the magnetization process and entanglement properties of the $\text{Ni}_4^{2+}\text{Gd}_4^{3+}$ complexes can be accurately captured by analyzing a single cubane unit \(\text{Ni}_2^{2+}\text{Gd}_2^{3+}\).
The study of entanglement between spin pairs, as well as among all spins in heterometallic mixed \(3d/4f\) complexes such as \(\text{Ni}_4^{2+}\text{Gd}_4^{3+}\), remains relatively unexplored due to the complexity of the underlying Hamiltonian.
Therefore, the main objective of this article is to investigate the tetrapartite entanglement negativity as a measure of total entanglement, the bipartite entanglement between each pair of spins, the 1--3 tangles between a single spin and the remaining three spins, and the 2--2 tangle between the subsystem comprising two Ni ions and the subsystem comprising two Gd ions within a single cubane \(\text{Ni}_2^{2+}\text{Gd}_2^{3+}\) unit.


The paper is structured as follows: In the next section, we provide a detailed description of the magnetic structure of $\text{Ni}_4^\text{2+}\text{Gd}_4^\text{3+}$ complexes and introduce the corresponding effective Hamiltonian. Section \ref{sec:results} presents the ground-state phase diagram, zero- and low-temperature magnetization process of these complexes, comparing key theoretical predictions with experimental data from two specific compounds: {\bf (1)} \( [\text{Ni}_4\text{Gd}_4(\text{HL})_4(\mu_2\text{-OH})_2(\mu_3\text{-OH})_4(\mu\text{-OOCCH}_3)_8] \cdot (\text{NO}_3) \cdot 12\text{H}_2\text{O} \) \cite{Biswas2020}, and {\bf (2)} \( [\text{Ni}_4\text{Gd}_4(\mu_2\text{-OH})_2(\mu_3\text{-OH})_4(\mu\text{-OOCCH}_3)_8(\text{LH}_2)_4] \) \cite{Kalita2018}. Hereafter, for simplicity, we refer to these compounds as {\bf (1)} and {\bf (2)}. Additionally, we examine how different forms of quantum entanglement manifest in these complexes under varying temperature and magnetic field conditions. Finally, Section \ref{conclusions} summarizes our key findings and conclusions. 

\section{Model}\label{Model}

\begin{figure}
\begin{center}
\resizebox{0.5\textwidth}{!}{%
 \includegraphics[trim=20 50 20 20, clip]{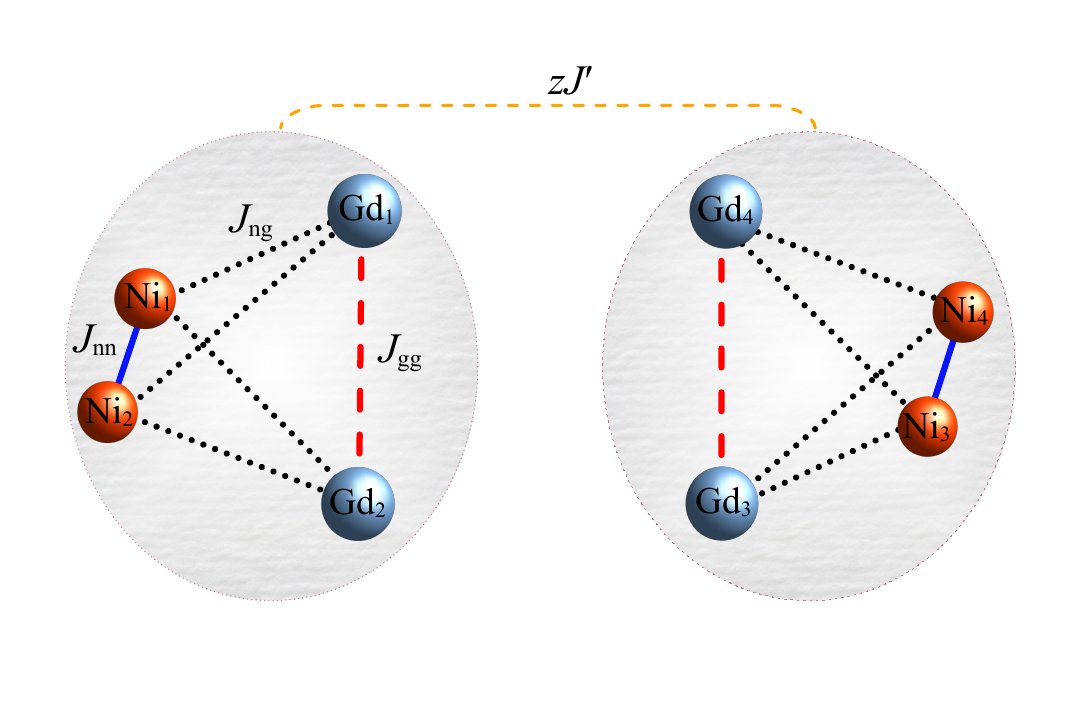}
}
\caption{Schematic representation of the molecular structure of $\text{Ni}_4^\text{2+}\text{Gd}_4^\text{3+}$ complexes. Orange spheres represent nickel ions, while light blue spheres denote gadolinium ions. The intra-cubane exchange interactions are defined as follows: blue solid lines indicate the exchange interaction $J_\text{nn}$ between Ni ions, black dotted lines represent the exchange interaction $J_\text{ng}$ between Ni and Gd ions, and red dashed lines correspond to the interaction $J_\text{gg}$ between Gd ions. The orange dashed line, labeled $zJ^{\prime}$, represents the inter-cubane interaction between the two identical $\text{Ni}_2^\text{2+}\text{Gd}_2^\text{3+}$ cubane units. }
\label{fig:figure1}
\end{center}
\end{figure}

Our motivation for studying the $\text{Ni}_4^\text{2+}\text{Gd}_4^\text{3+}$ complexes stems from previous experimental investigations of these systems \cite{Kalita2018, Biswas2020}. In this work, we consider the two interconnected heterometallic $\text{Ni}_2^\text{2+}\text{Gd}_2^\text{3+}\text{O}_4$ distorted cubane units as a single subsystem, as depicted in Fig. \ref{fig:figure1}.  
The Hamiltonian of the octanuclrear complexes  $\text{Ni}_4^\text{2+}\text{Gd}_4^\text{3+}$ can be modeled as
\begin{equation}\label{eq:hamiltonian}
	\begin{array}{lcl}
		H &=& J_\text{nn} \big[\hat{\boldsymbol{S}}_{\mathrm{Ni}_1} \cdot \hat{\boldsymbol{S}}_{\mathrm{Ni}_2} + \hat{\boldsymbol{S}}_{\mathrm{Ni}_3} \cdot \hat{\boldsymbol{S}}_{\mathrm{Ni}_4}\big] 
		\\[0.2cm]
		&& + J_\text{gg} \big[\hat{\boldsymbol{S}}_{\mathrm{Gd}_1} \cdot \hat{\boldsymbol{S}}_{\mathrm{Gd}_2} + 
		\hat{\boldsymbol{S}}_{\mathrm{Gd}_3} \cdot \hat{\boldsymbol{S}}_{\mathrm{Gd}_4}\big]
		\\[0.3cm]
		&& + J_\text{ng} \bigg[ \sum\limits_{\{a,b\}=1,2} \hat{\boldsymbol{S}}_{\mathrm{Ni}_a} \cdot \hat{\boldsymbol{S}}_{\mathrm{Gd}_b} + \sum\limits_{\{c,d\}=3,4} \hat{\boldsymbol{S}}_{\mathrm{Ni}_c} \cdot \hat{\boldsymbol{S}}_{\mathrm{Gd}_d} \bigg]
		\\[0.4cm]
		&& + D_\text{n} \sum\limits_{j=1}^4 \big( \hat{S}^z_{\mathrm{Ni}_j} \big)^2 
		+ zJ^{\prime} \langle \hat{S}^z_\mathrm{T} \rangle \hat{S}^z_\mathrm{T}
		\\[0.2cm]
		&& - \mu_B B \sum\limits_{a=1}^4 \big( g_\text{n} \hat{S}^z_{\mathrm{Ni}_a} + g_\text{g} \hat{S}^z_{\mathrm{Gd}_a} \big),
	\end{array}
\end{equation}
where $\hat{\boldsymbol{S}}$ indicate spatial effective components of the standard spin operators of the metal centre(s) Ni$^\text{2+}$ ($S = 1$) and Gd$^\text{3+}$ ($S = \frac{7}{2}$).
Here, ${J}_\text{nn}$, ${J}_\text{gg}$ and ${J}_\text{ng}$ are the exchange interactions between each ion(s) pair, namely  $\text{Ni}\cdots\text{Ni}$, 
$\text{Gd}\cdots\text{Gd}$ and $\text{Ni}\cdots\text{Gd}$, respectively. 
$zJ^{\prime}$ accounts for the inter-cubane interaction using the molecular mean-field approach.
The presence of low-lying excited states has been disclosed in such complexes \cite{Kalita2018,Biswas2020} that is in agreement with the existence of not strong $\text{Ni}\cdots\text{Gd}$ magnetic interactions. 
$B$ denotes the external magnetic field along the $z-$direction and  $g_\text{n}, g_\text{g}$ are g-factors of Ni and Gd ions. $D_\text{n}$ represents the single-ion anisotropy of the Ni ions along the local $z$-axis.
It should be noted that, as reported in Refs. \cite{Kalita2018,Biswas2020}, the value of single-ion anisotropy of $\text{Ni}^\text{2+}$ ions $D_\text{n}$ in the complexes {\bf (1)} and {\bf (2)} naturally depend on their geometry in the complexes. Hence, we consider a wide range of single-ion anisotropy and Ni$\cdots$Ni exchange interaction then examine the effects of these parameters on the quantum entanglement of the complexes.
The partition function \( Z \) of the considered compounds can be directly obtained by diagonalizing the Hamiltonian (\ref{eq:hamiltonian}). From the partition function, the Gibbs free energy is given by  
\(
G = -k_B T \ln Z
\) 
where \( k_B \) is the Boltzmann constant and \( T \) is the temperature. The magnetization of the system can then be determined using the thermodynamic relation 
\(
M = - \left( \frac{\partial G}{\partial B} \right)_T.
\)  

\section{Results and discussion}\label{sec:results}
Let us now discuss the most significant findings related to the pattern of spin levels, zero- and low-temperature magnetization, as well as the degree of various forms of quantum entanglement in compounds~\textbf{(1)} and~\textbf{(2)}.
These octanuclear heterometallic $3d/4f$ complexes consist of two interconnected \(\text{Ni}_2^\text{2+}\text{Gd}_2^\text{3+}\) subunits, linked via hydroxide and acetate bridges.
In accordance with previous experimental studies~\cite{Kalita2018, Biswas2020}, we assume that the inter-cubane interaction mediated by these bridges is very weak, with an estimated exchange coupling in the range \(-0.002 < zJ^{\prime} < 0.002\;\text{cm}^{-1}\).

\subsection{Magnetic properties}
\begin{figure*}
\includegraphics[scale=0.35,trim=10 0 20 50, clip]{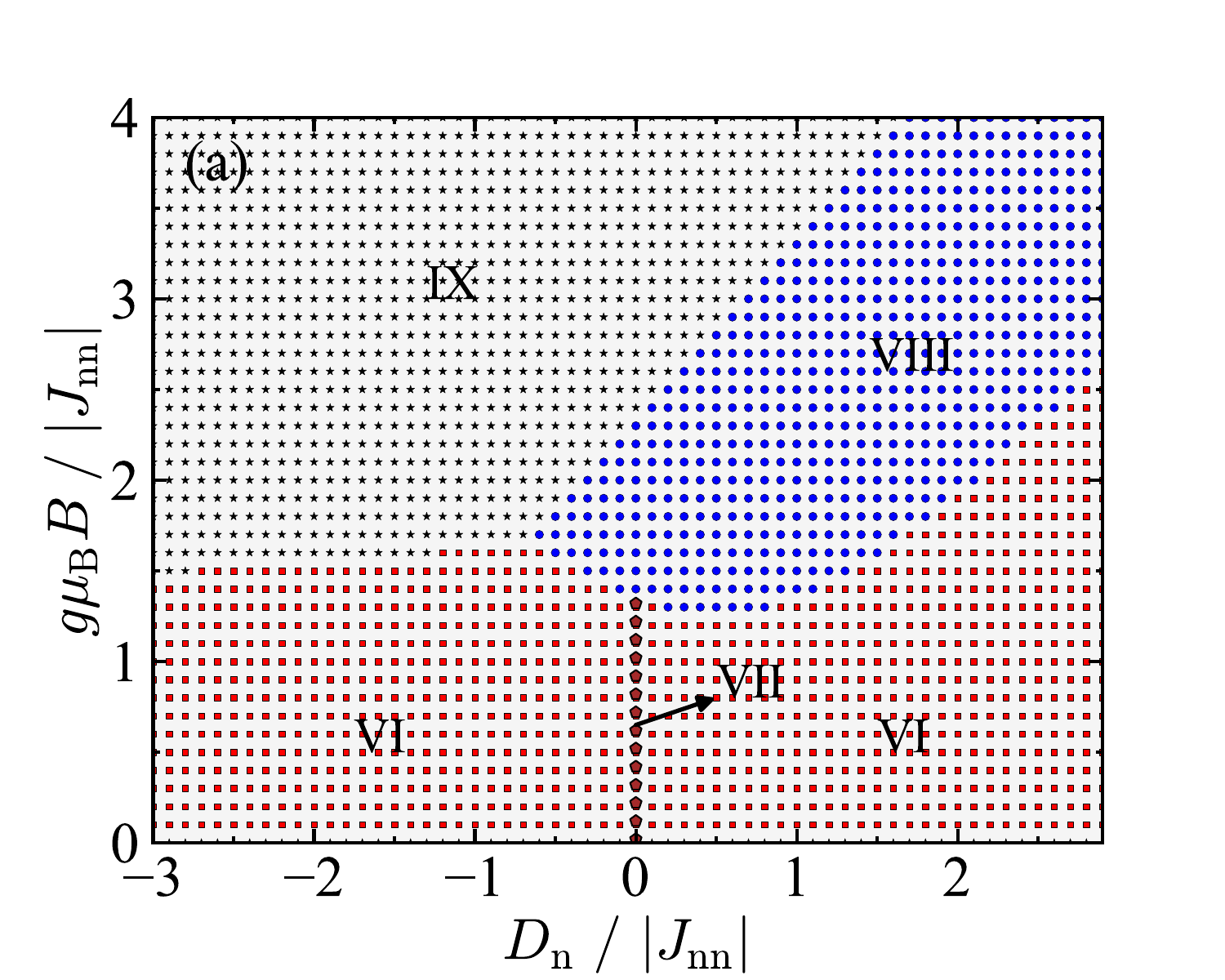}
\includegraphics[scale=0.35,trim=0 0 20 50, clip]{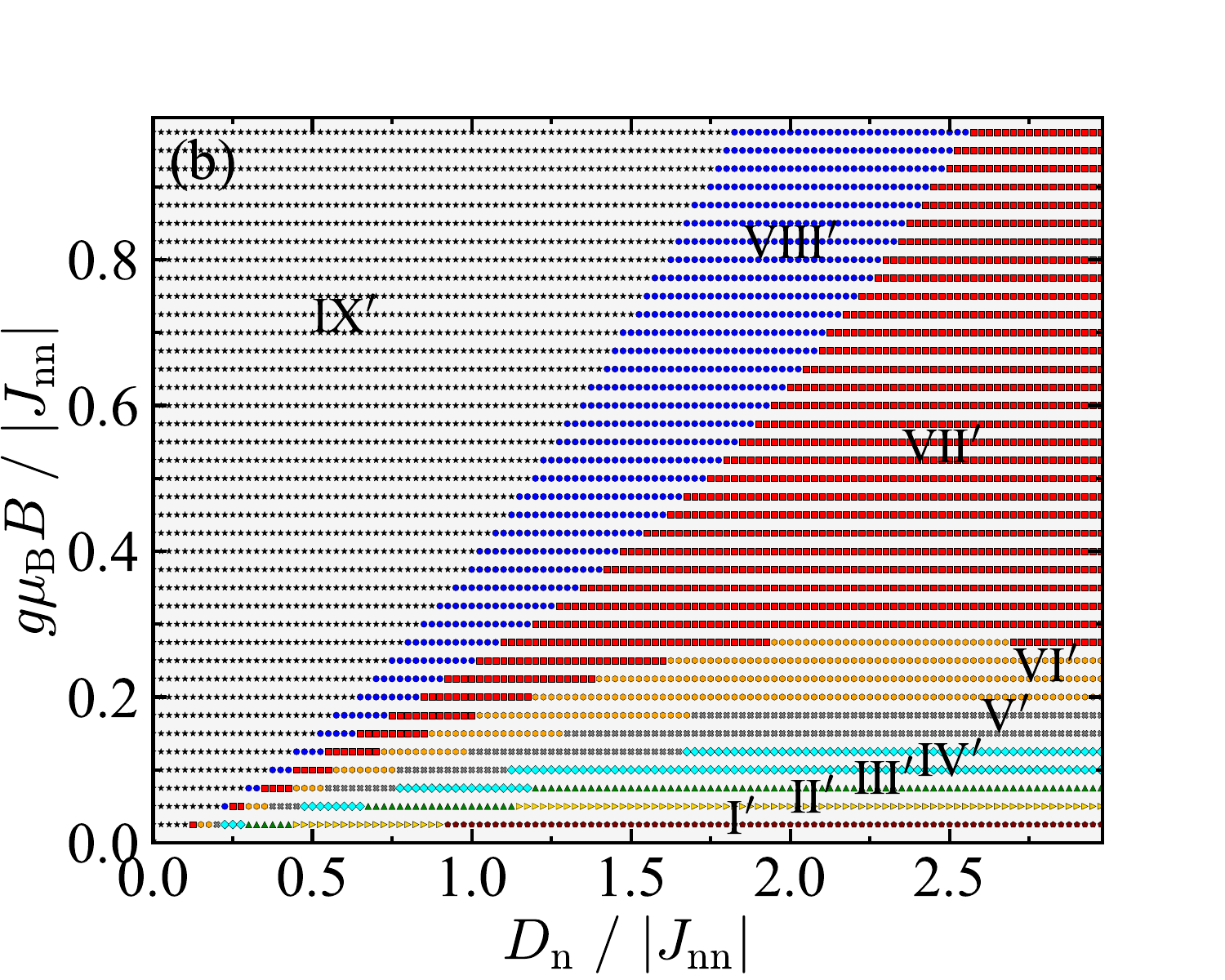} %
\includegraphics[scale=0.35,trim=10 10 20 50, clip]{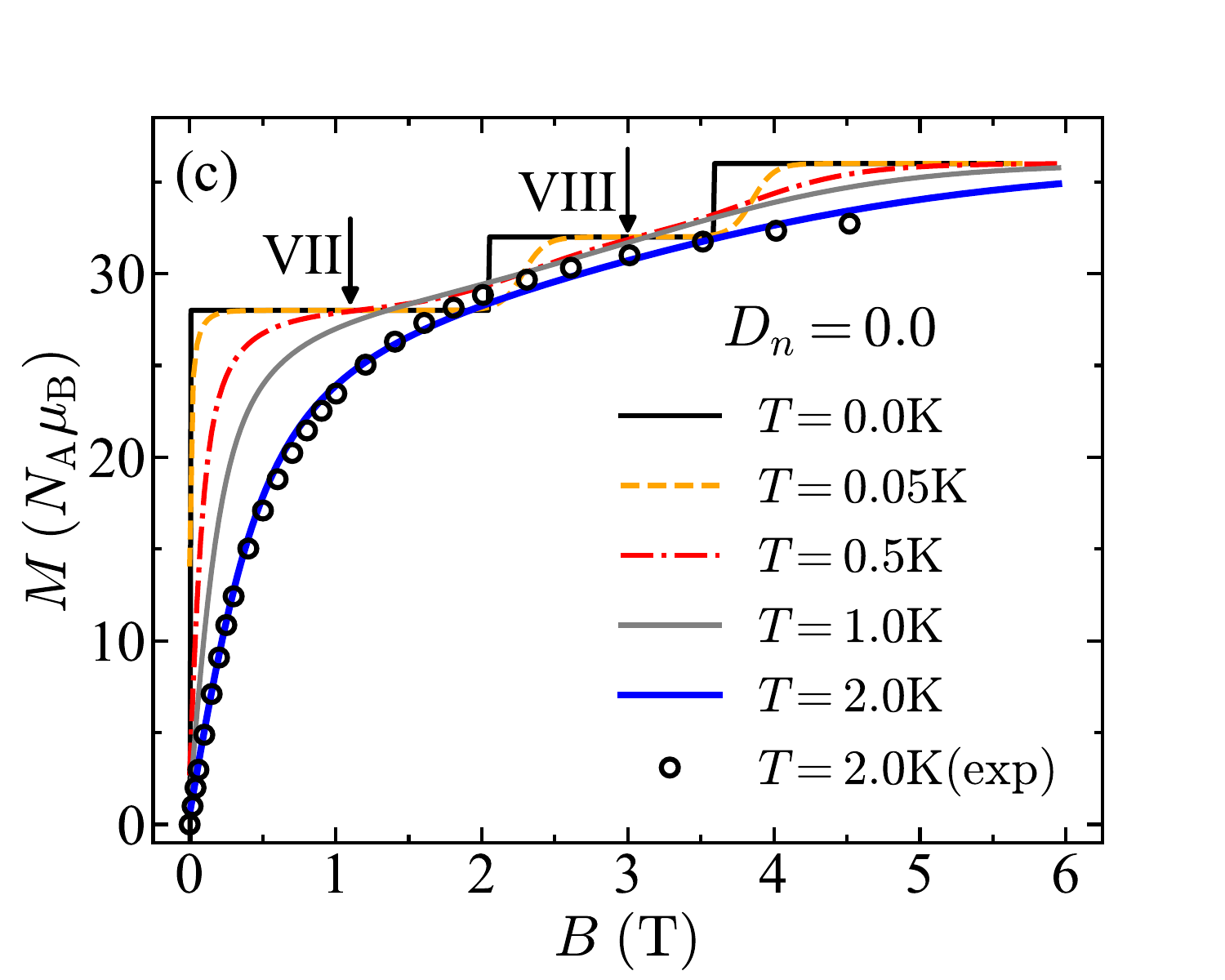}
\includegraphics[scale=0.35,trim=10 10 20 50, clip]{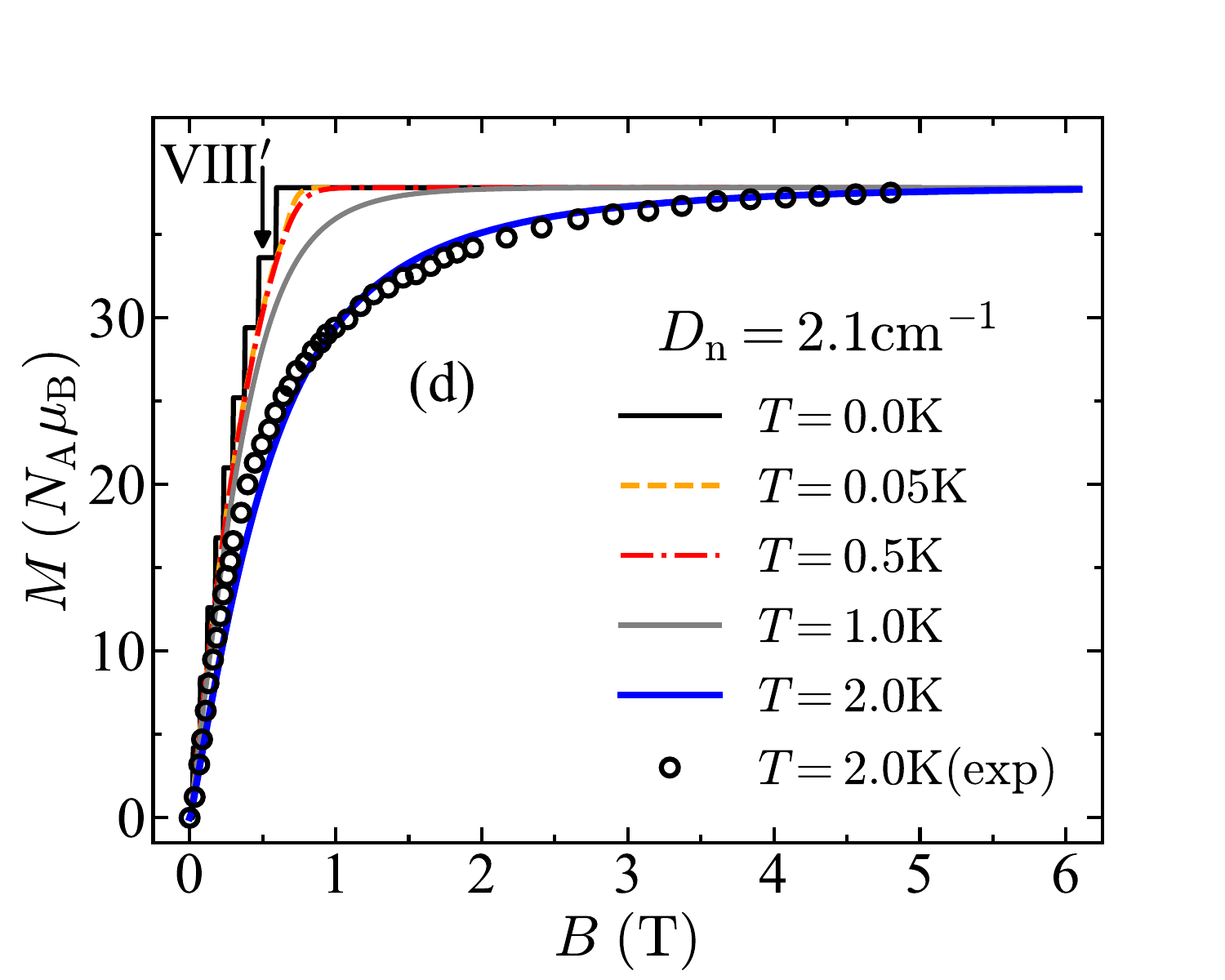}
\caption{
(a) The ground-state phase diagram of the \( \text{Ni}_4^\text{2+}\text{Gd}_4^\text{3+} \) complexes in the \((g\mu_\text{B}B/|J_\text{nn}|, D_\text{n}/|J_\text{nn}|)\)-plane, assuming antiferromagnetic \( J_\text{nn} > 0 \) for the parameter set: \( J_\text{ng}/|J_\text{nn}| = 0.05 \) and \( J_\text{gg}/|J_\text{nn}| = -0.2 \).  
(b) The ground-state phase diagram of the compounds, assuming ferromagnetic \( J_\text{nn} < 0 \) for the parameter set: \( J_\text{ng}/|J_\text{nn}| = -0.16 \) and \( J_\text{gg}/|J_\text{nn}| = 0.001 \).  
Each symbol indicates a specific ground state with unique magnetization value. In panels (a) and (b), the phase regions are labeled using Roman numerals corresponding to the magnetization values as described in the text.   
(c) Black circles represent the experimental magnetization data of the complex {\bf (1)} from Ref. \cite{Biswas2020}. The blue solid line shows the exact numerical results for the magnetization, obtained using the parameter set: \( T = 2\,\text{K} \), \( D_\text{n} = 0 \), \( J_\text{nn} = 1.53\;\text{cm}^{-1} \), \( J_\text{ng} = 0.0074\;\text{cm}^{-1} \), \( J_\text{gg} = -0.288\;\text{cm}^{-1} \), $g_\text{n} = g_\text{g} = 2.0$ and  inter-cubane interaction \( zJ^{\prime} = 0.0013\;\text{cm}^{-1} \), achieving a good fit with experimental data. Other curves correspond to magnetization at lower temperatures. 
(d) Black circles represent the experimental magnetization data of the complex {\bf (2)} from Ref. \cite{Kalita2018}. The blue solid line shows the exact numerical results for the magnetization, obtained using the parameter set: \( T = 2\,\text{K} \), \( D_\text{n} = 2.1 \), \( J_\text{nn} = -5.2\;\text{cm}^{-1} \), \( J_\text{ng} = -0.86\;\text{cm}^{-1} \), \( J_\text{gg} = 0.0034\;\text{cm}^{-1} \), $g_\text{n} = g_\text{g} = 2.1$ and \( zJ^{\prime} = -0.0002\;\text{cm}^{-1} \). Other curves correspond to magnetization at lower temperatures.}
\label{fig:MagExp_ED}
\end{figure*}

In the following, we analyze in detail the magnetization process of the compounds $\text{Ni}_4^\text{2+}\text{Gd}_4^\text{3+}$ with Hamiltonian (\ref{eq:hamiltonian}), which includes the tunable single-ion anisotropy term \( D_\text{n} \), in the presence of an external uniform magnetic field \( B \).  
To gain insight into how the spin ground states of the \( \text{Ni}_4^\text{2+}\text{Gd}_4^\text{3+} \) complexes manifest, we examine the zero-temperature magnetization for two different sets of the Hamiltonian's parameters.  
In Fig.~\ref{fig:MagExp_ED}(a), the ground-state phase diagram of the theoretical model described by Hamiltonian (\ref{eq:hamiltonian}), in the \( (g\mu_\text{B}B/|J_\text{nn}|, D_\text{n}/|J_\text{nn}|) \)-plane is shown.
We here assume an antiferromagnetic exchange interaction \( J_\text{ng}/|J_\text{nn}| = 0.05 \) and a ferromagnetic exchange interaction \( J_\text{gg}/|J_\text{nn}| = -0.2 \) with \( J_\text{nn} > 0 \) taken as the energy unit.  
The phase diagram in Fig. \ref{fig:MagExp_ED}(a) identifies larger regions associated to VI, VII, VIII, and IX,  
each corresponding to distinct ground states with magnetization values of $\frac{7}{9}$ and $\frac{8}{9}$, and full polarization, respectively.  The spin configuration of these states is given by:
\begin{subequations}\label{eq:VIVIIVIII_1}
	\begin{align}
		|\text{VI}\rangle &= 
		- \text{c}_1 \left( \left| +1, +1, +\tfrac{7}{2}, +\tfrac{3}{2} \right\rangle 
		+ \left| +1, +1, +\tfrac{3}{2}, +\tfrac{7}{2} \right\rangle \right) \nonumber\\
		&\quad - \text{c}_2 \left| +1, +1, +\tfrac{5}{2}, +\tfrac{5}{2} \right\rangle \nonumber\\
		&\quad + \text{c}_3 \big( 
		\left| +1, 0, +\tfrac{7}{2}, +\tfrac{5}{2} \right\rangle + 
		\left| +1, 0, +\tfrac{5}{2}, +\tfrac{7}{2} \right\rangle \nonumber\\
		&\qquad\quad + \left| 0, +1, +\tfrac{7}{2}, +\tfrac{5}{2} \right\rangle + 
		\left| 0, +1, +\tfrac{5}{2}, +\tfrac{7}{2} \right\rangle \big) \nonumber\\
		&\quad + \text{c}_4 \left( 
		\left| +1, -1, +\tfrac{7}{2}, +\tfrac{7}{2} \right\rangle + 
		\left| -1, +1, +\tfrac{7}{2}, +\tfrac{7}{2} \right\rangle \right) \nonumber\\
		&\quad - \text{c}_5 \left| 0, 0, +\tfrac{7}{2}, +\tfrac{7}{2} \right\rangle,
		\label{eq:VI} \\[6pt]
		|\text{VII}\rangle &= 
		\frac{1}{\sqrt{3}} \big( 
		\left| +1, -1, +\tfrac{7}{2}, +\tfrac{7}{2} \right\rangle 
		- \left| 0, 0, +\tfrac{7}{2}, +\tfrac{7}{2} \right\rangle \nonumber\\
		&\qquad + \left| -1, +1, +\tfrac{7}{2}, +\tfrac{7}{2} \right\rangle \big),
		\label{eq:VII} \\[6pt]
		|\text{VIII}\rangle &= 
		\frac{1}{\sqrt{2}} \left( 
		\left| +1, 0, +\tfrac{7}{2}, +\tfrac{7}{2} \right\rangle 
		- \left| 0, +1, +\tfrac{7}{2}, +\tfrac{7}{2} \right\rangle \right).
		\label{eq:VIII}
	\end{align}
\end{subequations}
State~VI hosts two magnetic excitations so called magnons, as the kets on the right-hand side of Eq.~(\ref{eq:VI}) can be generated by the action of the lowering operators $\hat{S}^{-}_i \hat{S}^{-}_j$ on the $i$th and $j$th ions of the ferromagnetic state~IX.
The vertical line of pentagon symbols shown in Fig.~\ref{fig:MagExp_ED}(a) corresponds to state VII, where the two magnons are localized on the Ni$\cdots$Ni bond. As a matter of fact, state~VII originates from the dispersive two-magnon state~VI by imposing $\text{c}_1 = \text{c}_2 = \text{c}_3 = 0$ and $\text{c}_4 = \text{c}_5 = \tfrac{1}{\sqrt{3}}$ at $D_\text{n}/|J_\text{nn}| = 0$. In this case, the two magnons remain confined to the Ni$\cdots$Ni bond and Ni ions are decoupled from Gd ions. However, when $D_\text{n}/|J_\text{nn}| \neq 0$, the magnetic excitations are found throughout the cubane unit, giving rise to state~VI. As the single-ion anisotropy increases, the region corresponding to phase~VIII (with $M/M_\text{s} = \tfrac{8}{9}$) expands accordingly.

On the other hand, 
 in Fig.~\ref{fig:MagExp_ED}(b), we examine a different parameter set: a ferromagnetic interaction \( J_\text{ng}/|J_\text{nn}| = -0.16 \), a weak antiferromagnetic interaction \( J_\text{gg}/|J_\text{nn}| = 0.001 \), and a ferromagnetic exchange coupling \( J_\text{nn} < 0 \), which serves as the energy unit. Under these conditions, for $D_\text{n}\neq 0$ the system exhibits distinct ground states I$^{\prime}$, II$^{\prime}$, III$^{\prime}$, IV$^{\prime}$, V$^{\prime}$, VI$^{\prime}$, VII$^{\prime}$, VIII$^{\prime}$, and IX$^{\prime}$, each characterized by unique magnetization values at  
\( M/M_\text{s} = \frac{1}{9}, \frac{2}{9}, \frac{1}{3}, \frac{4}{9}, \frac{5}{9}, \frac{2}{3}, \frac{7}{9}, \frac{8}{9} \), and full saturation. 
The spin arrangement of the W-like state VIII$^{\prime}$ with intermediate $\frac{8}{9}$ magnetization plateau (see Fig. \ref{fig:MagExp_ED}(d)), hosts one magnon and can be presented by:
\begin{eqnarray}\label{eq:VIIVIII_2}
	&& |\text{VIII}^{\prime}\rangle =\ \alpha_1 \left( \left| +1, 0, +\tfrac{7}{2}, +\tfrac{7}{2} \right\rangle 
	+ \left| 0, +1, +\tfrac{7}{2}, +\tfrac{7}{2} \right\rangle \right) \nonumber\\
	&& \quad + \alpha_2 \left( \left| +1, +1, +\tfrac{7}{2}, +\tfrac{5}{2} \right\rangle 
	+ \left| +1, +1, +\tfrac{5}{2}, +\tfrac{7}{2} \right\rangle \right).
\end{eqnarray}
In the above equations, the coefficients $\text{c}_i$, $\alpha_1$, and $\alpha_2$ are lengthy polynomials that depend on the Hamiltonian parameters and will be computed numerically in the following sections.
The choice of these two specific parameter sets and representing the exact spin configuration of the eigenvectors (\ref{eq:VIVIIVIII_1})and (\ref{eq:VIIVIII_2})  will contribute to experimental realizations of the theoretical model corresponding to the complexes {\bf (1)} and {\bf (2)}, which will be discussed in what follows.
 
In Fig.~\ref{fig:MagExp_ED}(c), we compare our theoretical magnetization results at \( T = 2\,\text{K} \) (blue solid line) with experimental data of the complex {\bf (1)} from Ref.~\cite{Biswas2020}, represented by black circles. A good fit with the experimental magnetization data was obtained by assuming \( D_\text{n} = 0 \) and selecting the exchange interaction parameters:  
 \( J_\text{nn} = 1.53\;\text{cm}^{-1} \), \( J_\text{ng} = 0.0074\;\text{cm}^{-1} \), \( J_\text{gg} = -0.288\;\text{cm}^{-1} \), $g_\text{n} = g_\text{g} = 2.0$ and  inter-cubane interaction \( zJ^{\prime} = 0.0013\;\text{cm}^{-1} \) as reported in Ref.~\cite{Biswas2020}.  
At zero temperature (black solid line), the magnetization exhibits an abrupt jump from zero field to an intermediate plateau at \( M/M_\text{s} = \frac{7}{9} \) associated to the state VII (\ref{eq:VII}). This is followed by a second transition to the narrower plateau at \( M/M_\text{s} = \frac{8}{9} \) corresponding to the state VIII (\ref{eq:VIII}) around \( B \approx 2.1\,\text{T} \), before finally reaching saturation at \( B \approx 3.6\,\text{T} \). As the temperature increases, these intermediate magnetization plateaus gradually disappear.  

Figure~\ref{fig:MagExp_ED}(d) presents the magnetization behavior  of the complex {\bf (2)} reported in Ref.~\cite{Kalita2018} at zero and different nonzero temperatures, assuming a nonzero single-ion anisotropy \( D_\text{n} = 2.1\,\text{cm}^{-1}\), ferromagnetic interactions \( J_\text{nn} = -5.2\;\text{cm}^{-1}, \quad J_\text{ng} = -0.86\;\text{cm}^{-1}\), \(J_\text{gg} = 0.0034\;\text{cm}^{-1}\), $g_\text{n} = g_\text{g} = 2.1$ along with a very weak inter-cubane exchange interaction   \( zJ^{\prime} = -0.0002\;\text{cm}^{-1} \).   
Under these circumstances, at zero temperature, the magnetization manifests several intermediate plateaus in accordance with the ground states shown in Fig.~\ref{fig:MagExp_ED}(b). At low temperature (dashed line), the magnetization follows a distinct pattern as a function of the applied magnetic field. Specifically, within the interval \( 0 < B < 1.0\,\text{T} \), the magnetization increases rapidly with the applied field before stabilizing at a very narrow plateau \( M/M_\text{s} = \frac{8}{9} \) corresponds to the ground state VIII$^{\prime}$ (\ref{eq:VIIVIII_2}) with normalized coefficients $\alpha_1 = 0.411$ and $\alpha_2 = 0.575$, then reaches saturation magnetization at \( B\approx 0.6\,\text{T} \).  

\subsection{{ Spin-Level Structure}}

To elucidate the microscopic origin of the low-temperature magnetic and entanglement properties, we first analyze the spin-level spectrum derived from the isotropic part of the Hamiltonian (Eq.~\ref{eq:hamiltonian}), then compare the results with those obtained from the Hamiltonian comprising magnetic field and single-ion anisotropy. isotropic part of the Hamiltonian includes only the Heisenberg-type exchange interactions:
\begin{equation}
	\begin{array}{lcl}
		&& {H}_\text{iso} = J_\text{nn} \big[\hat{\boldsymbol{S}}_{\mathrm{Ni}_1} \cdot \hat{\boldsymbol{S}}_{\mathrm{Ni}_2} + \hat{\boldsymbol{S}}_{\mathrm{Ni}_3} \cdot \hat{\boldsymbol{S}}_{\mathrm{Ni}_4}\big] 
		\\[0.2cm]
		&& + J_\text{gg} \big[\hat{\boldsymbol{S}}_{\mathrm{Gd}_1} \cdot \hat{\boldsymbol{S}}_{\mathrm{Gd}_2} + 
		\hat{\boldsymbol{S}}_{\mathrm{Gd}_3} \cdot \hat{\boldsymbol{S}}_{\mathrm{Gd}_4}\big]
		\\[0.3cm]
		&& + J_\text{ng} \bigg[ \sum\limits_{\{a,b\}=1,2} \hat{\boldsymbol{S}}_{\mathrm{Ni}_a} \cdot \hat{\boldsymbol{S}}_{\mathrm{Gd}_b} + \sum\limits_{\{c,d\}=3,4} \hat{\boldsymbol{S}}_{\mathrm{Ni}_c} \cdot \hat{\boldsymbol{S}}_{\mathrm{Gd}_d} \bigg].
		\end{array}
\end{equation}
This Hamiltonian commutes with the total spin projection operator \( \hat{S}_\text{T}^z \), which ensures the conservation of total magnetization and allows for block-diagonalization of the full Hilbert space (dimension 576) into subspaces labeled by \( S_\text{T}^z \). By employing standard spin-coupling and irreducible tensor operator techniques~\cite{Schnack2008,Schnack2010,Heitmann2019}, the computational complexity is significantly reduced while simultaneously revealing the underlying structure of the energy spectrum.

\begin{figure*}
	\includegraphics[scale=0.28,trim=10 0 60 50, clip]{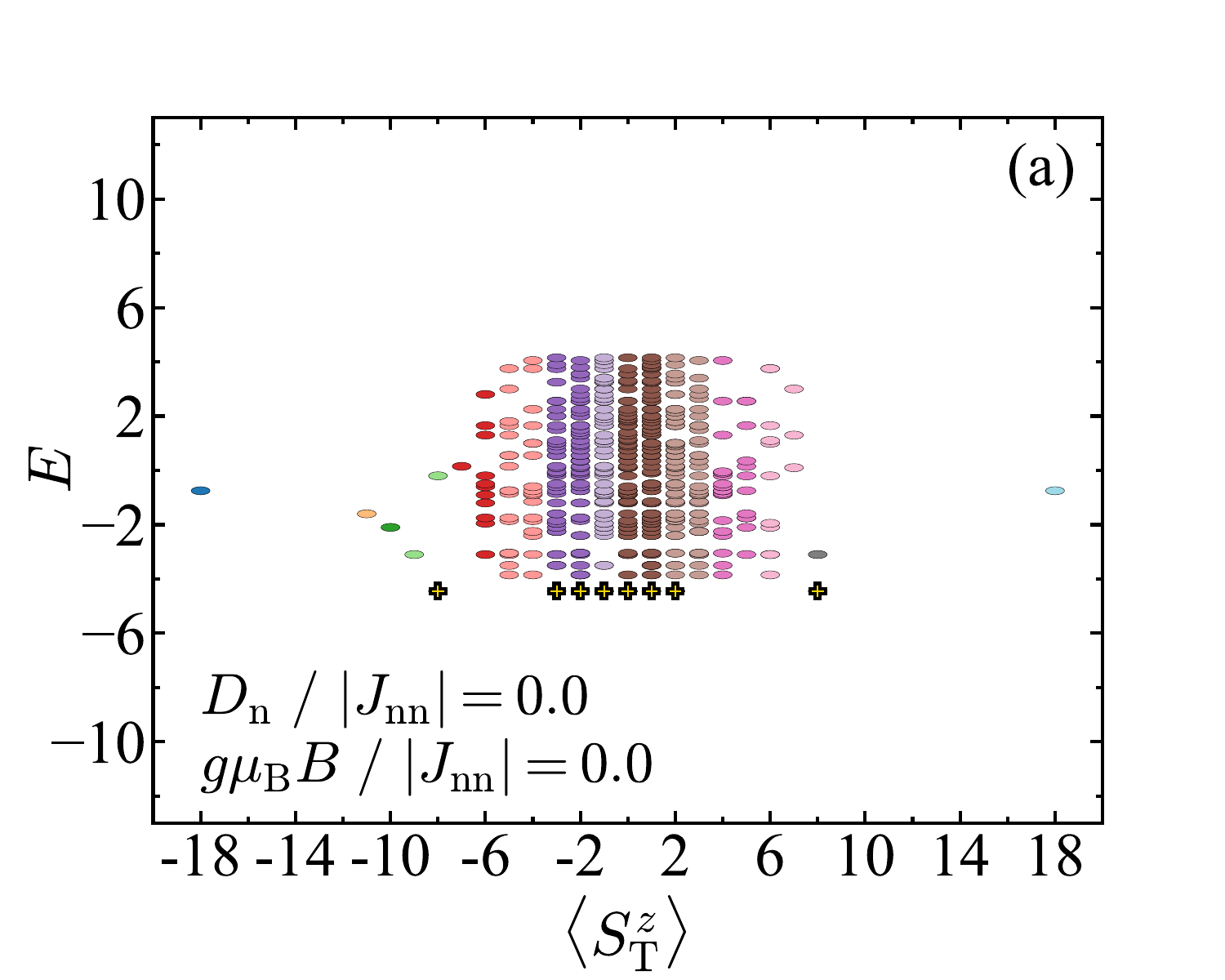}
	\includegraphics[scale=0.28,trim=40 0 60 50, clip]{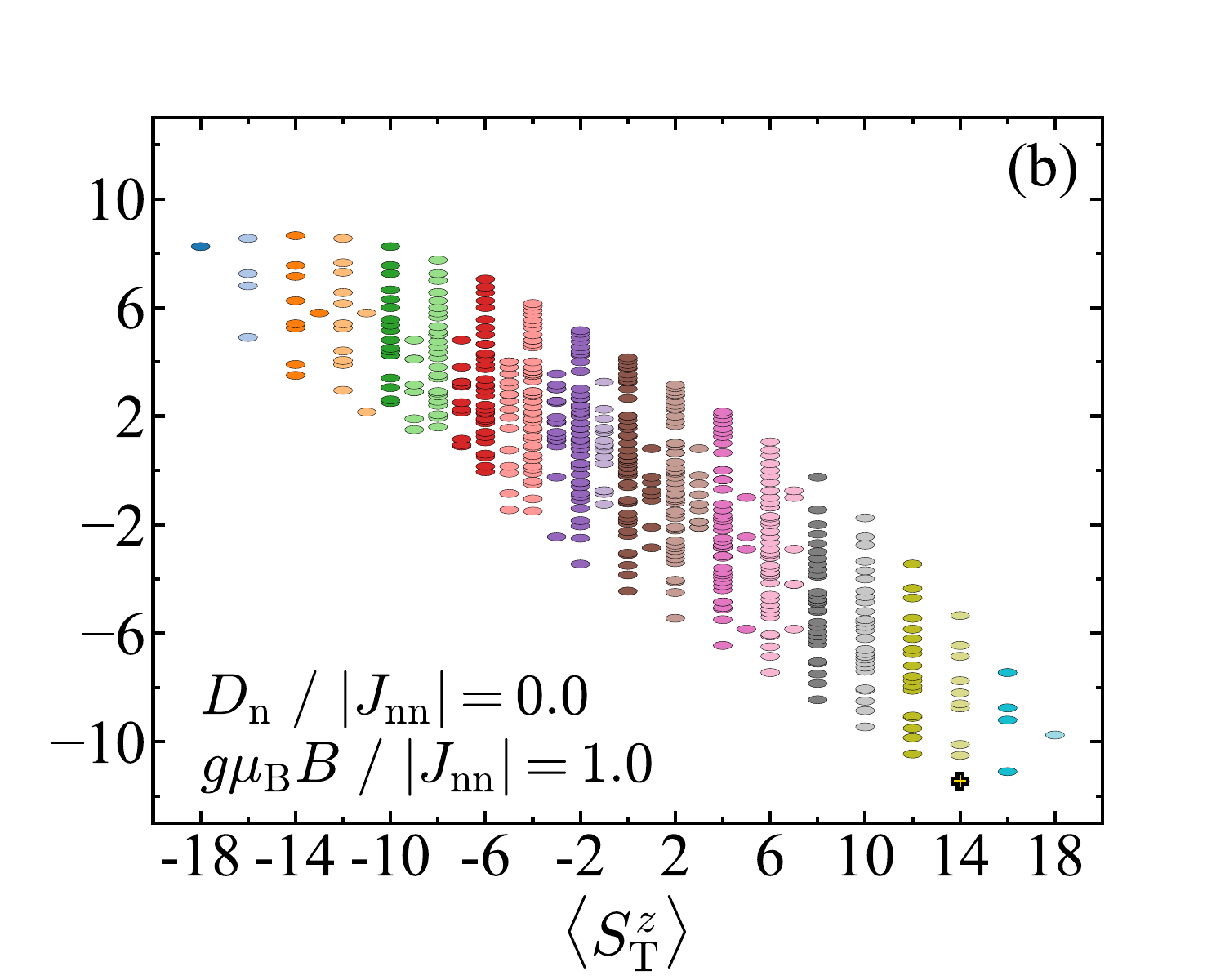} %
	\includegraphics[scale=0.28,trim=10 0 60 50, clip]{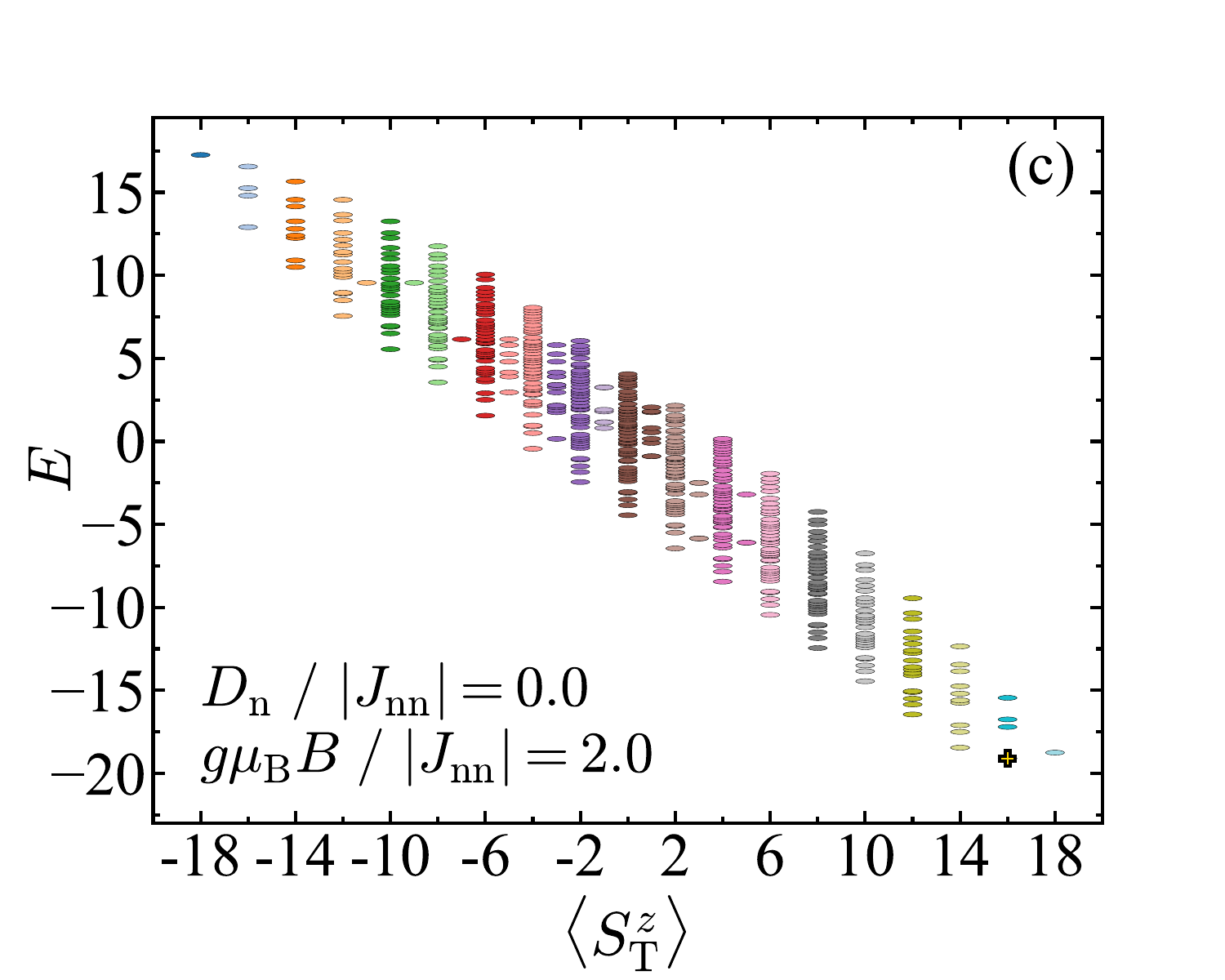}
	\includegraphics[scale=0.28,trim=40 0 60 50, clip]{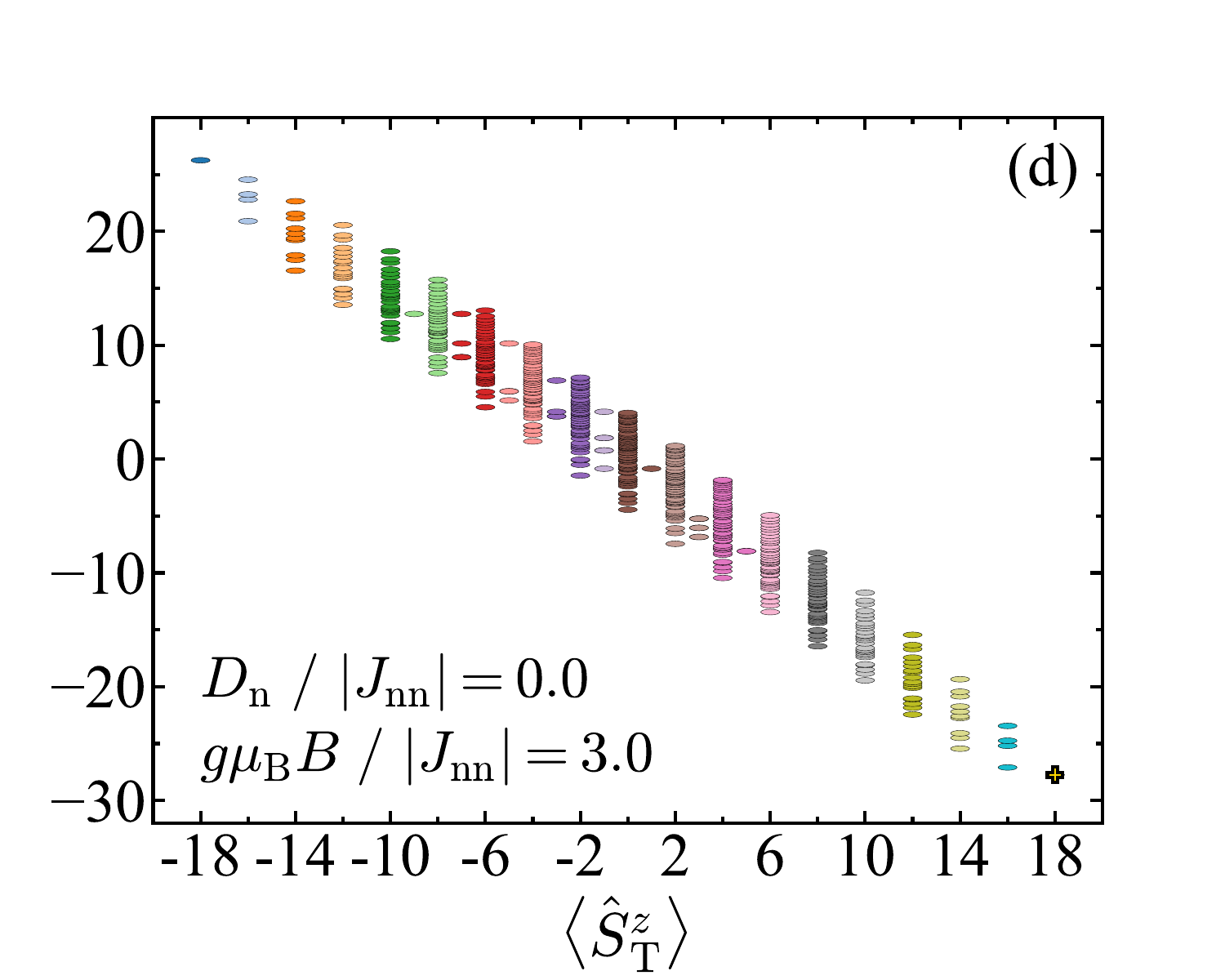}
	\caption{
		The energy levels of the \( \text{Ni}_4^{2+}\text{Gd}_4^{3+} \) complex as a function of the total magnetization \( \langle \hat{S}_\text{T}^z \rangle \), assuming an antiferromagnetic exchange \( J_\text{nn} > 0 \). The parameter set used is: \( J_\text{ng}/|J_\text{nn}| = 0.05 \), \( J_\text{gg}/|J_\text{nn}| = -0.2 \), and \( D_\text{n}/|J_\text{nn}| = 0 \).  
		Panels correspond to increasing external magnetic field:  
		(a) \( g\mu_\text{B}B/|J_\text{nn}| = 0 \),  
		(b) \( g\mu_\text{B}B/|J_\text{nn}| = 1 \),  
		(c) \( g\mu_\text{B}B/|J_\text{nn}| = 2 \),  
		(d) \( g\mu_\text{B}B/|J_\text{nn}| = 3 \).  
		Ellipses represent the eigenenergies in each magnetization sector, and crosses mark the ground state energy within each \( \langle \hat{S}_\text{T}^z \rangle \) subspace.
	}
	\label{fig:spinlevels_1}
\end{figure*}

Figure~\ref{fig:spinlevels_1} shows the energy eigenvalues (ellipses) plotted against their corresponding mean spin \( \langle \hat{S}_\text{T}^z \rangle \), for a representative parameter set: \( J_\text{ng}/|J_\text{nn}| = 0.05 \), \( J_\text{gg}/|J_\text{nn}| = -0.2 \), and \( D_\text{n}/|J_\text{nn}| = 0 \). These points correspond to the vertical line of pentagons in Fig.~\ref{fig:MagExp_ED}(a). In zero field (Fig.~\ref{fig:spinlevels_1}(a)), the spectrum exhibits symmetric degeneracies across magnetization sectors that may originate from the nature of exchange couplings.
The lowest-energy levels, marked by golden crosses are   degenerate and
depends sensitively on the signs and relative magnitudes of the exchange couplings \( J_{\mathrm{nn}}, J_{\mathrm{ng}}, \) and \( J_{\mathrm{gg}} \).
 
As the magnetic field increases (Figs.~\ref{fig:spinlevels_1}(b)--\ref{fig:spinlevels_1}(d)), the Zeeman interaction lifts the degeneracies and shifts the ground state toward higher \( \langle \hat{S}_\text{T}^z \rangle \). In Fig.~\ref{fig:spinlevels_1}(b), the ground state corresponds to VII with \( \langle \hat{S}_\text{T}^z \rangle = 14 \). With further increasing of the field, the ground state continues to evolve, as shown in Figs.~\ref{fig:spinlevels_1}(c) and \ref{fig:spinlevels_1}(d). The progression of golden crosses across the panels highlights the field-induced reconfiguration of the ground state. Introducing a finite anisotropy \( D_n \neq 0 \) will further lift degeneracies and alter the ground state structure, which will be explored in detail in the next discussion concerning the complex~\textbf{(2)}.

\begin{figure*}
	\includegraphics[scale=0.26,trim=10 0 60 50, clip]{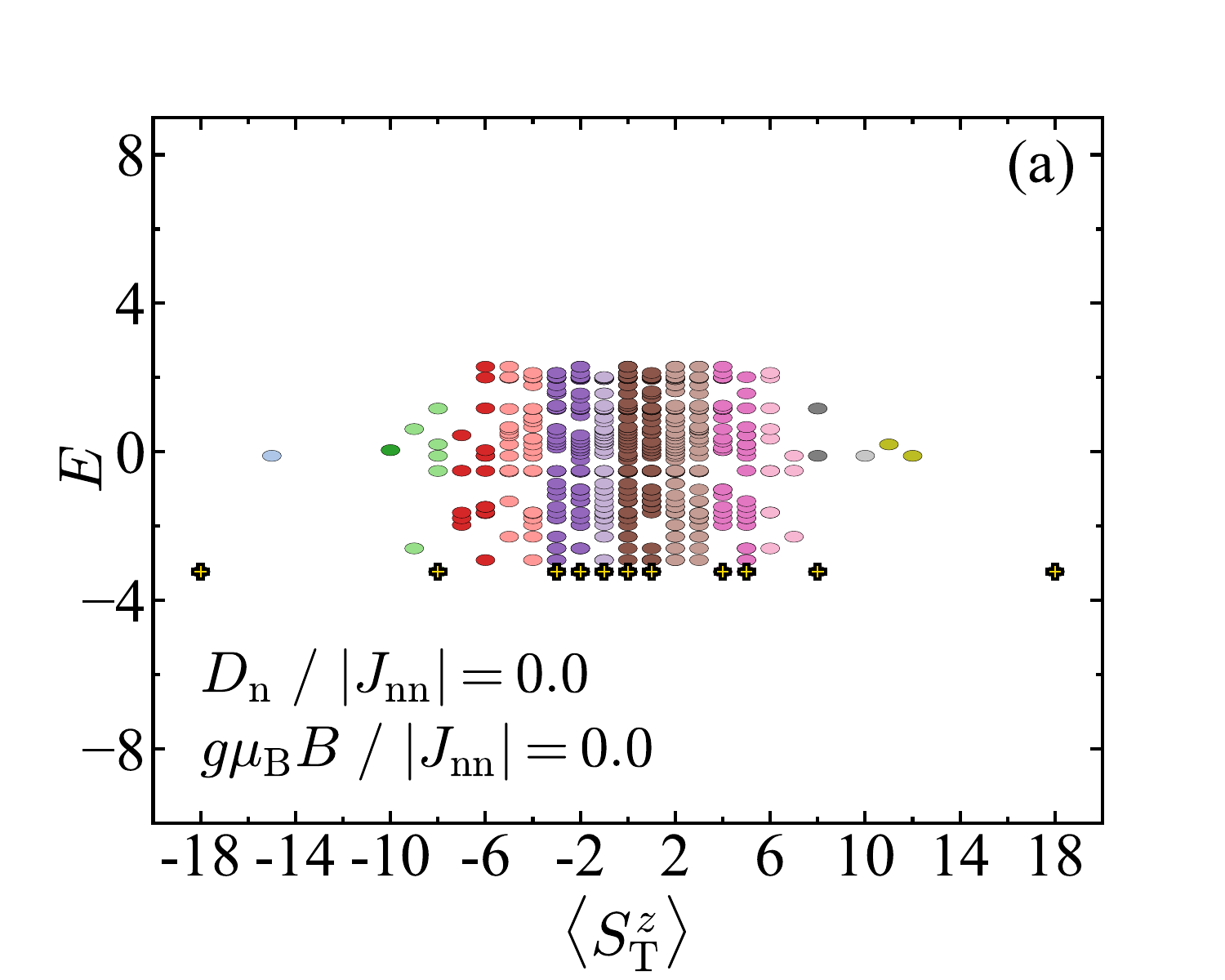}
	\includegraphics[scale=0.26,trim=40 0 60 50, clip]{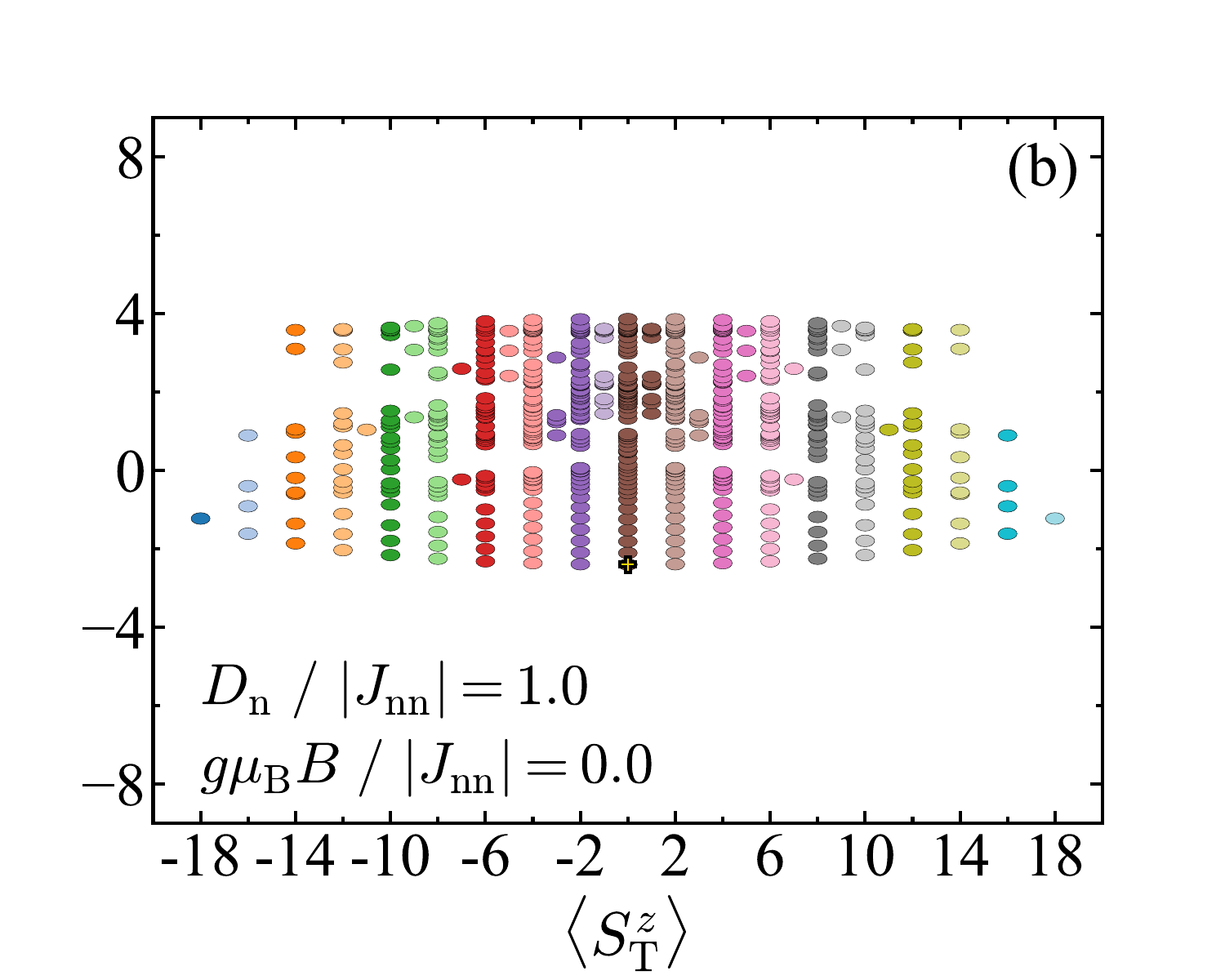} %
	\includegraphics[scale=0.26,trim=40 0 60 50, clip]{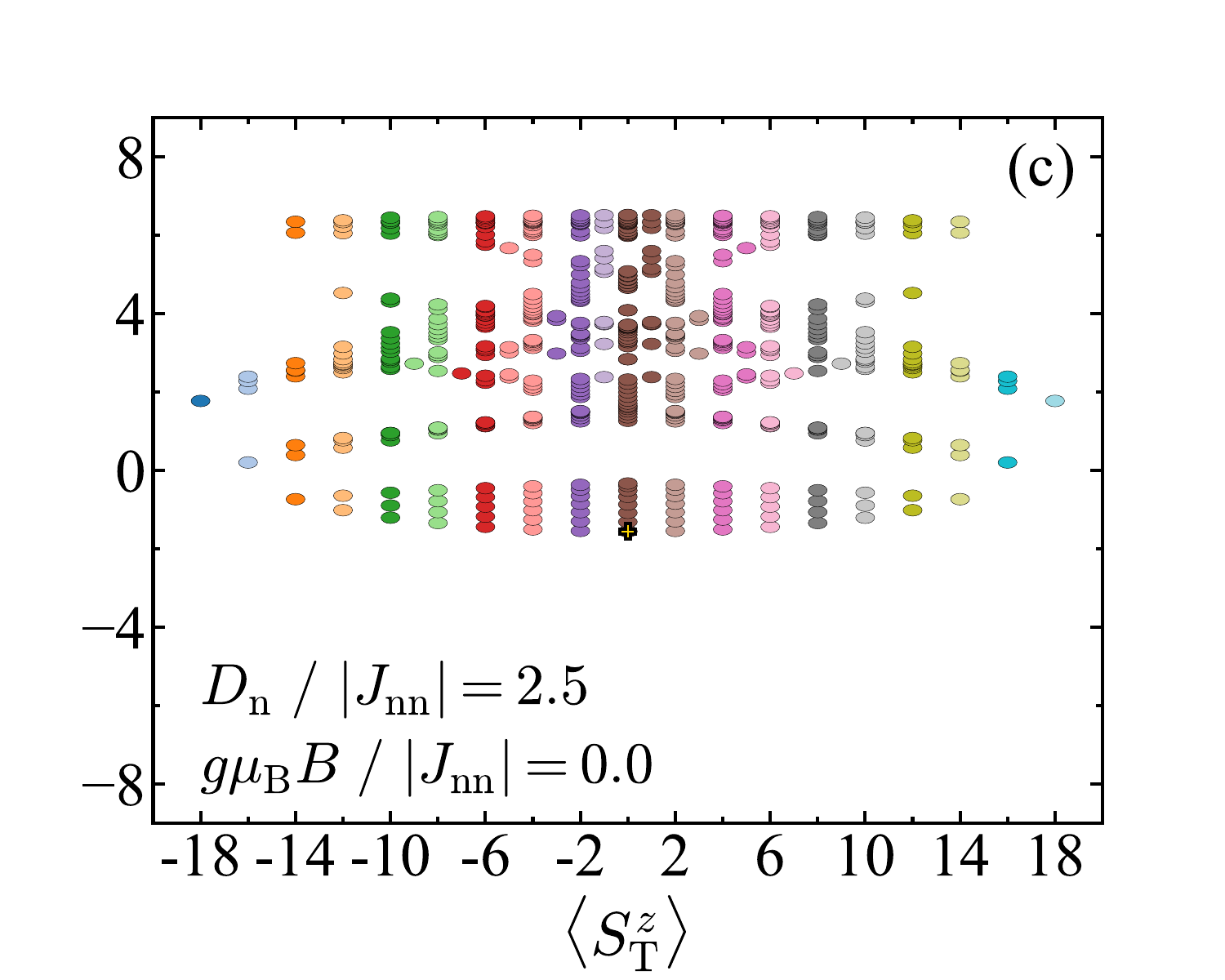}
	\includegraphics[scale=0.26,trim=10 0 60 50, clip]{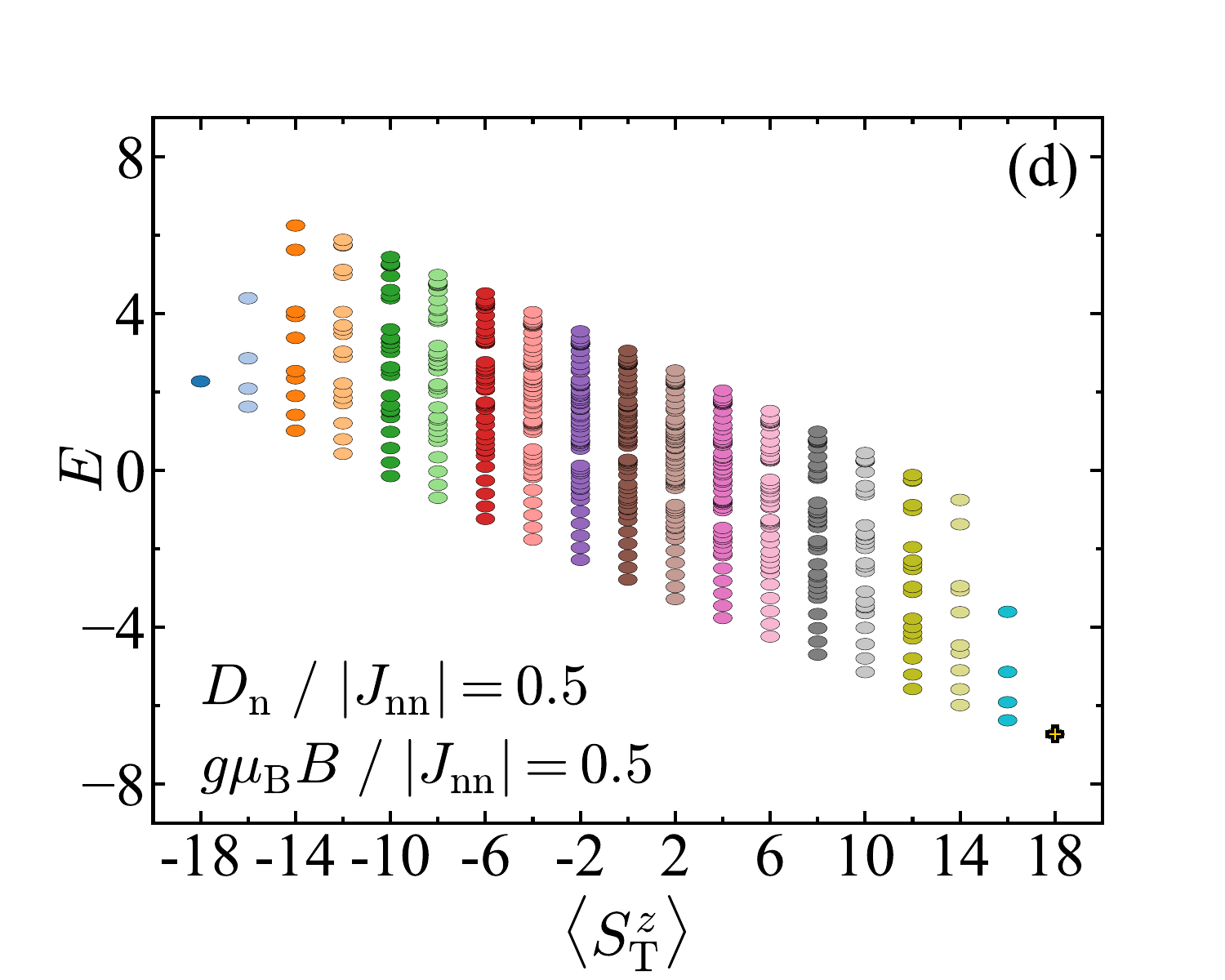}
	\includegraphics[scale=0.26,trim=40 0 60 50, clip]{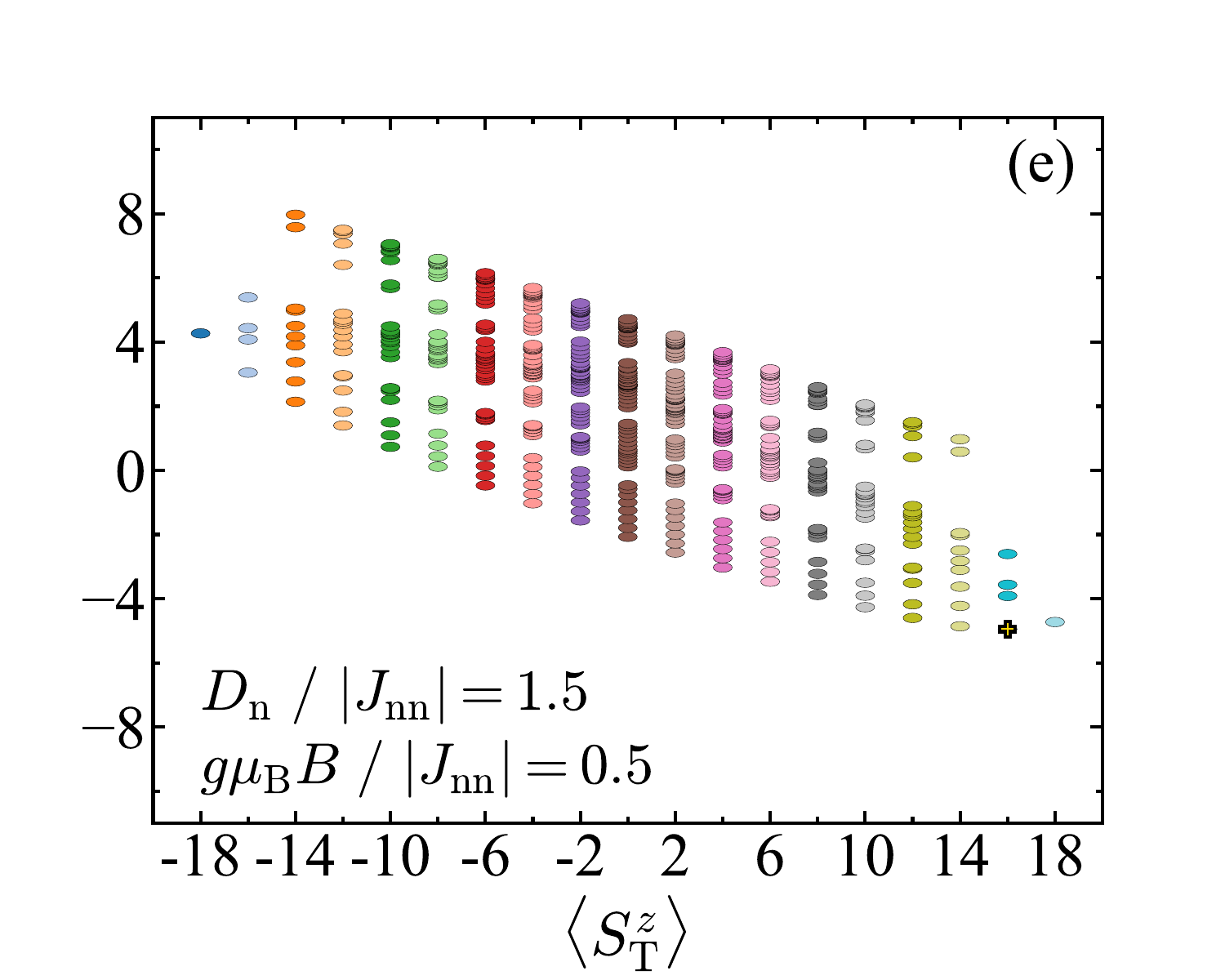}
	\includegraphics[scale=0.26,trim=40 0 60 50, clip]{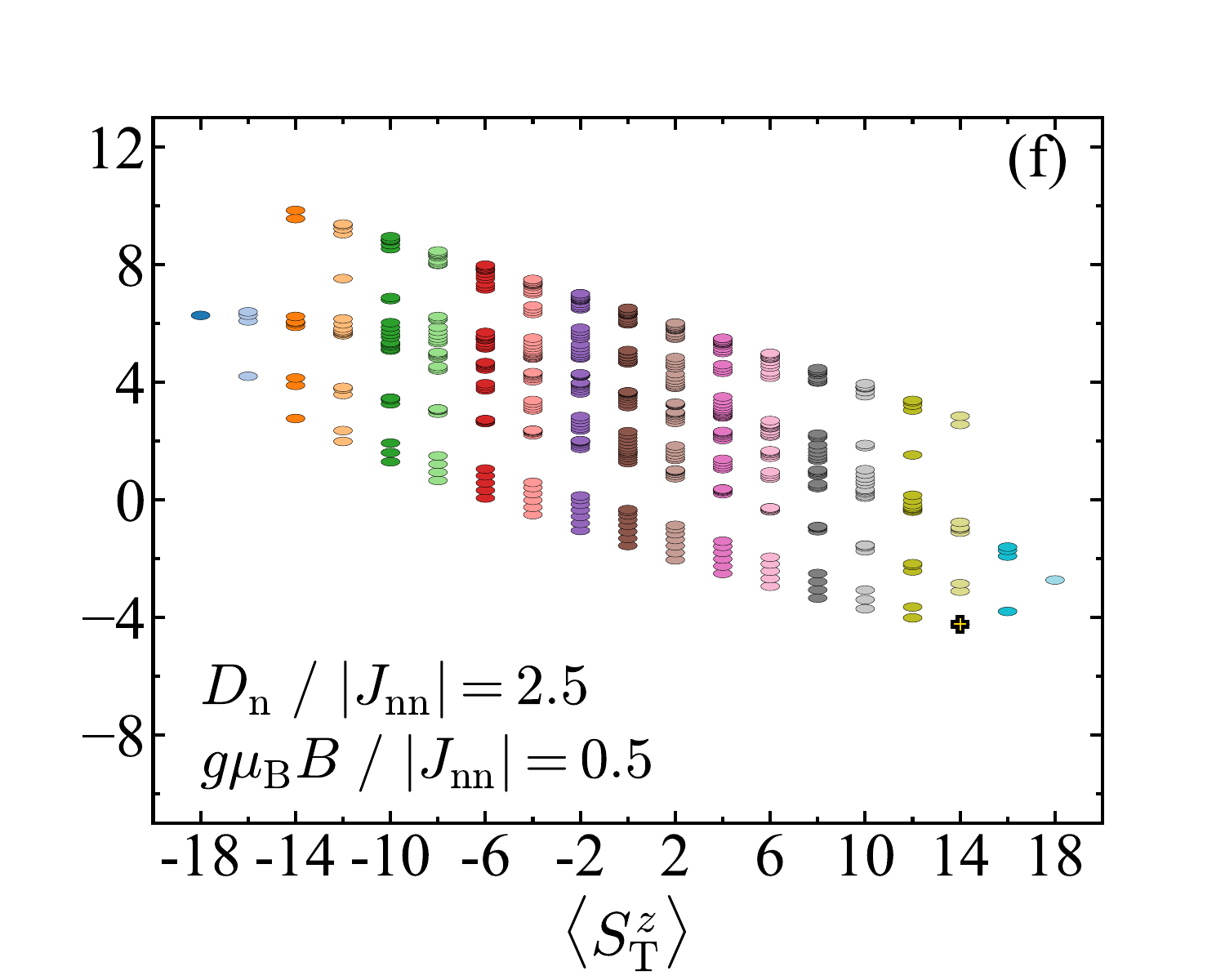}
	\caption{
		The energy levels of the \( \text{Ni}_4^{2+}\text{Gd}_4^{3+} \) complex as a function of the total magnetization \( \langle \hat{S}_\text{T}^z \rangle \), assuming  ferromagnetic \( J_\text{nn} < 0 \) and the parameter set: \( J_\text{ng}/|J_\text{nn}| = -0.16 \) and \( J_\text{gg}/|J_\text{nn}| = 0.001 \).  
		Panels correspond to increasing external magnetic field and single-ion anisotropy:  
		(a) \( g\mu_\text{B}B/|J_\text{nn}| =  D_\text{n}/|J_\text{nn}| = 0 \),  
		(b) \( g\mu_\text{B}B/|J_\text{nn}| = 0 \) and \( D_\text{n}/|J_\text{nn}| = 1.0 \),  
		(c) \( g\mu_\text{B}B/|J_\text{nn}| = 0 \) and \( D_\text{n}/|J_\text{nn}| = 2.5 \),  
		(d) \( g\mu_\text{B}B/|J_\text{nn}| =  D_\text{n}/|J_\text{nn}| = 0.5 \),
		(e) \( g\mu_\text{B}B/|J_\text{nn}| = 0.5 \) and \( D_\text{n}/|J_\text{nn}| = 1.5 \),
		(f) \( g\mu_\text{B}B/|J_\text{nn}| = 0.5 \) and \( D_\text{n}/|J_\text{nn}| = 2.5 \).  
	}
	\label{fig:spinlevels_2}
\end{figure*}

Figure~\ref{fig:spinlevels_2} illustrates the evolution of the energy spectrum of the \( \text{Ni}_4^{2+}\text{Gd}_4^{3+} \) complex as a function of  \( \langle \hat{S}_\text{T}^z \rangle \), for a ferromagnetic coupling between Ni sites (\( J_\text{nn} < 0 \)) and the parameter set \( J_\text{ng}/|J_\text{nn}| = -0.16 \), \( J_\text{gg}/|J_\text{nn}| = 0.001 \). The panels systematically show the impact of increasing both the single-ion anisotropy \( D_\text{n} \) and the external magnetic field \( B \) on the structure and degeneracies of the spin energy levels.
In the absence of both anisotropy and external field (see Fig. \ref{fig:spinlevels_2}(a)) the system exhibits highly degenerate energy levels.
Under this conditions, the ground state corresponds to a high-spin multiplet indicated by golden crosses, which mark the lowest-energy eigenstate within each \( S^z_\text{T} \) sector. 
Introducing finite anisotropy \( D_\text{n} \) while keeping \( B = 0 \) (Figs. \ref{fig:spinlevels_2}(b) and \ref{fig:spinlevels_2}(c)) lifts many of these degeneracies and begins to deform the multiplet structure. As the anisotropy strength increases from \( D_\text{n}/|J_\text{nn}| = 1.0 \) to \( 2.5 \), the spectrum becomes more anisotropic, and the symmetry between positive and negative magnetization sectors becomes stronger with respect to \( \langle \hat{S}_\text{T}^z \rangle = 0 \), reflecting full spin-rotational symmetry. The lifting of degeneracy and reordering of states result in a broader energy spread, with low-lying levels still centered near zero magnetization.

When both magnetic field and anisotropy are non-zero (see Figs. \ref{fig:spinlevels_2}(d)--\ref{fig:spinlevels_2}(f)), the Zeeman effect further lifts degeneracies and shifts the energy levels toward higher \( \langle \hat{S}_\text{T}^z \rangle \) values that favors spin alignment with the field. Panel~\ref{fig:spinlevels_2}(d) shows the case of moderate field and anisotropy (\( g\mu_\text{B}B/|J_\text{nn}| = D_\text{n}/|J_\text{nn}| = 0.5 \)), where partial lifting of degeneracy occurs, while the ground state energy corresponds to the fully polarized state IX$^{\prime}$ with \( \langle \hat{S}_\text{T}^z \rangle =18\) (see Fig. \ref{fig:MagExp_ED}(b)). As shown in Figs. \ref{fig:spinlevels_2}(e) and \ref{fig:spinlevels_2}(f), when the anisotropy increases further, the spectrum becomes more structured, and the ground state migrates to sectors with large positive magnetization indicating ground states VIII$^{\prime}$ and VII$^{\prime}$ (see Fig. \ref{fig:MagExp_ED}(b)). These changes reflect a competition between the Zeeman energy and single-ion anisotropy, and the position of the golden crosses tracks the field-induced crossover in the ground state magnetization.
Overall, this figure demonstrates how anisotropy and magnetic field together control the energetic accessibility and symmetry of spin states in the \( \text{Ni}_4^{2+}\text{Gd}_4^{3+} \) complex. These factors will be key in determining low-temperature magnetic and quantum behavior of such molecular systems.


\subsection{Thermal entanglement}
The degree of entanglement in a system including spin $S>1/2$ can be quantified based on the number of negative eigenvalues of its partial transpose, using measures of negativity, which serve as entanglement monotones under general positive partial transpose (PPT) preserving operations \cite{Peres1996}. 
To extend the definition of the tetrapartite entanglement in the single cubane unit of complexes {\bf (1)} and {\bf (2)}, we introduce the whole entanglement measure \( {\Pi}_4 \) \cite{Arian2022}. 
For a given tetrapartie cubane unit $\text{Ni}_2^\text{2+}\text{Gd}_2^\text{3+}$, the following negativities can be described:
\begin{subequations}\label{eq:negativities}
	\begin{align}
		\mathcal{N}_\text{n1n2} &= \dfrac{\|\rho ^{\text{T}_\text{n1}}_\text{n1n2}(T)\| -1}{2} = \sum_{i,\lambda_i<0}|\lambda_i|, \label{negativities_a} 
		\end{align}
		\begin{align}
		\mathcal{N}_\text{g1g2} &= \dfrac{\|\rho ^{\text{T}_\text{g1}}_\text{g1g2}(T)\| -1}{2} = \sum_{i,\eta_i<0}|\eta_i|, \label{negativities_b} \\
		\mathcal{N}_\text{n1g1} &= \dfrac{\|\rho ^{\text{T}_\text{n1}}_\text{n1g1}(T)\| -1}{2} = \sum_{i,\xi_i<0}|\xi_i|, \label{negativities_c} \\
		\mathcal{N}_\text{n1(n2g1g2)} &= \dfrac{\|\rho ^{\text{T}_\text{n1}}_\text{n1n2g1g2}(T)\| -1}{2} = \sum_{i,\Lambda_i<0}|\Lambda_i|, \label{negativities_d} \\
		\mathcal{N}_\text{g1(n1n2g2)} &= \dfrac{\|\rho ^{\text{T}_\text{g1}}_\text{n1n2g1g2}(T)\| -1}{2} = \sum_{i,\Omega_i<0}|\Omega_i|, \label{negativities_e}\\
		\mathcal{N}_\text{n1n2(g1g2)} &= \dfrac{\|\rho ^{\text{T}_\text{n1n2}}_\text{n1n2g1g2}(T)\| -1}{2} = \sum_{i,\Gamma_i<0}|\Gamma_i|, \label{negativities_f}
	\end{align}
\end{subequations}
in which \( \rho_\text{n1n2g1g2}(T)\) denotes the reduced density matrix of the cubane unit in thermal equilibrium at temperature \( T \).
We note that, due to the specific properties of the compounds studied in this work, particularly the very weak inter-cubane interaction \( zJ' \) compared to the intra-cubane interactions, the total density matrix of the octanuclear heterometallic complex \(\text{Ni}_4^{2+}\text{Gd}_4^{3+}\) can be effectively expressed as a tensor product of the density matrices of two identical \(\text{Ni}_2^{2+}\text{Gd}_2^{3+}\) cubane subunits, i.e., 
\begin{equation}\label{ABCD}
\rho_\text{tot}(T) \equiv \rho_{\mathrm{n}_1\mathrm{n}_2\mathrm{g}_1\mathrm{g}_2}(T) \otimes \rho_{\mathrm{n}_3\mathrm{n}_4\mathrm{g}_3\mathrm{g}_4}(T).
\end{equation}
As a result, the entanglement properties of the full complex can be accurately analyzed by focusing on a single cubane unit, which we henceforth denote as \(\rho_{\mathrm{n}_1\mathrm{n}_2\mathrm{g}_1\mathrm{g}_2}(T)\).
Notations $\text{\{n1, n2, g1, g2\}}$ denote $\{\text{Ni}_1, \text{Ni}_2, \text{Gd}_1, \text{Gd}_2\}$. $ \lambda_i, \eta_i, \xi_i, \Lambda_i, \Omega_i$ and $\Gamma_i$ are negative eigenvalues of the reduced density matrices mentioned in Eqs. \ref{eq:negativities}(a)-\ref{eq:negativities}(f).
The terms \(\text{T}_\text{n1} \) and \(\text{T}_\text{n1n2} \) indicate the partial transpose over \( \text{n1} \) and pair \( \text{n1n2} \), respectively, and \( \vert\vert \cdot \vert\vert \) stands for the trace norm of a matrix, represented as \( \vert\vert O \vert\vert= \mathrm{Tr} {\sqrt {O^{\dagger}O}} \). Quantities
$\mathcal{N}_{\text{n1(n2g1g2)}}$ and  $\mathcal{N}_{\text{g1(n1n2g2)}}$ are 1--3 tangles,  $\mathcal{N}_{\text{n1n2(g1g2)}}$ is 2--2 tangle, and  $\mathcal{N}_{\text{n1n2}}$, $\mathcal{N}_{\text{n1g1}}$, $\mathcal{N}_{\text{g1g2}}$ are 1--1 tangles. For instance, \( \mathcal{N}_{\text{n1(n2g1g2)}} \) describes the entanglement between part \( \text{n1} \) and the others \( \{\text{n2},\text{g1},\text{g2}\} \). Similarly,
\( \mathcal{N}_{\text{n1n2(g1g2)}} \) represents the negativity between pair \( \{\text{n1},\text{n2}\} \) and the remaining pair \( \{\text{g1},\text{g2}\} \).
\( \mathcal{N}_{\text{n1n2}} \) introduces the bipartite negativity between \( \text{n1} \) and \( \text{n2} \), where \( \rho _{\text{n1n2}}=\mathrm{Tr}_{\text{g1g2}}[\rho_{\text{n1n2g1g2}}] \). \( \mathcal{N}_{\text{n1g1}} \) accounts for the bipartite negativity between \( \text{n1} \) and \( \text{g1} \), where \( \rho _{\text{n1g1}}=\mathrm{Tr}_{\text{n2g2}}[\rho_{\text{n1n2g1g2}}] \).  
Here, the \( 1-1 \) and \( 1-3 \) tangles satisfy the following Coffman–Kundu–Wootters (CKW) monogamy inequality relation \cite{Wootters2000}:
\begin{equation}\label{ABCD}
	\begin{array}{lcl}
		\mathcal{N}^2_{\text{n1(n2g1g2)}}\geq \mathcal{N}^2_{\text{n1n2}}+\mathcal{N}^2_{\text{n1g1}}+\mathcal{N}^2_{\text{n1g2}}.
	\end{array}
\end{equation}
Accordingly, the four-tangle entanglement (residual tangles) can be characterized as
\begin{equation}\label{abcd}
	\begin{array}{lcl}
		\pi_\text{n1}=\mathcal{N}^2_{\text{n1(n2g1g2)}}-\mathcal{N}^2_{\text{n1n2}}-\mathcal{N}^2_{\text{n1g1}}-\mathcal{N}^2_{\text{n1g2}},
		\\
		\pi_\text{n2}=\mathcal{N}^2_{\text{n2(n1g1g2)}}-\mathcal{N}^2_{\text{n2n1}}-\mathcal{N}^2_{\text{n2g1}}-\mathcal{N}^2_{\text{n2g2}},
		\\
		\pi_\text{g1}=\mathcal{N}^2_{\text{g1(n1n2g2)}}-\mathcal{N}^2_{\text{g1n1}}-\mathcal{N}^2_{\text{g1n2}}-\mathcal{N}^2_\text{g1g2},
		\\
		\pi_\text{g2}=\mathcal{N}^2_\text{g2(n1n2g1)}-\mathcal{N}^2_\text{g2n1}-\mathcal{N}^2_\text{g2n2}-\mathcal{N}^2_\text{g2g1}.
	\end{array}
\end{equation}
In this work we will examine the tetrapartite entanglement in a cubane unit using the geometric mean \( {\Pi}_4 \) \cite{Arian2022}, that is given by
\begin{equation}\label{CD}
	\begin{array}{lcl}
		{\Pi}_4=\sqrt[4]{\pi_\text{n1}\pi_\text{n2}\pi_\text{g1}\pi_\text{g2}}.
	\end{array}
\end{equation}
The reduced density matrices, their transpositions, and corresponding eigenvalues are computed using precise numerical techniques implemented in the {\it QuTip} package \cite{Johansson2012}. 

\begin{figure*}
	\includegraphics[scale=0.28,trim=20 100 10 80, clip]{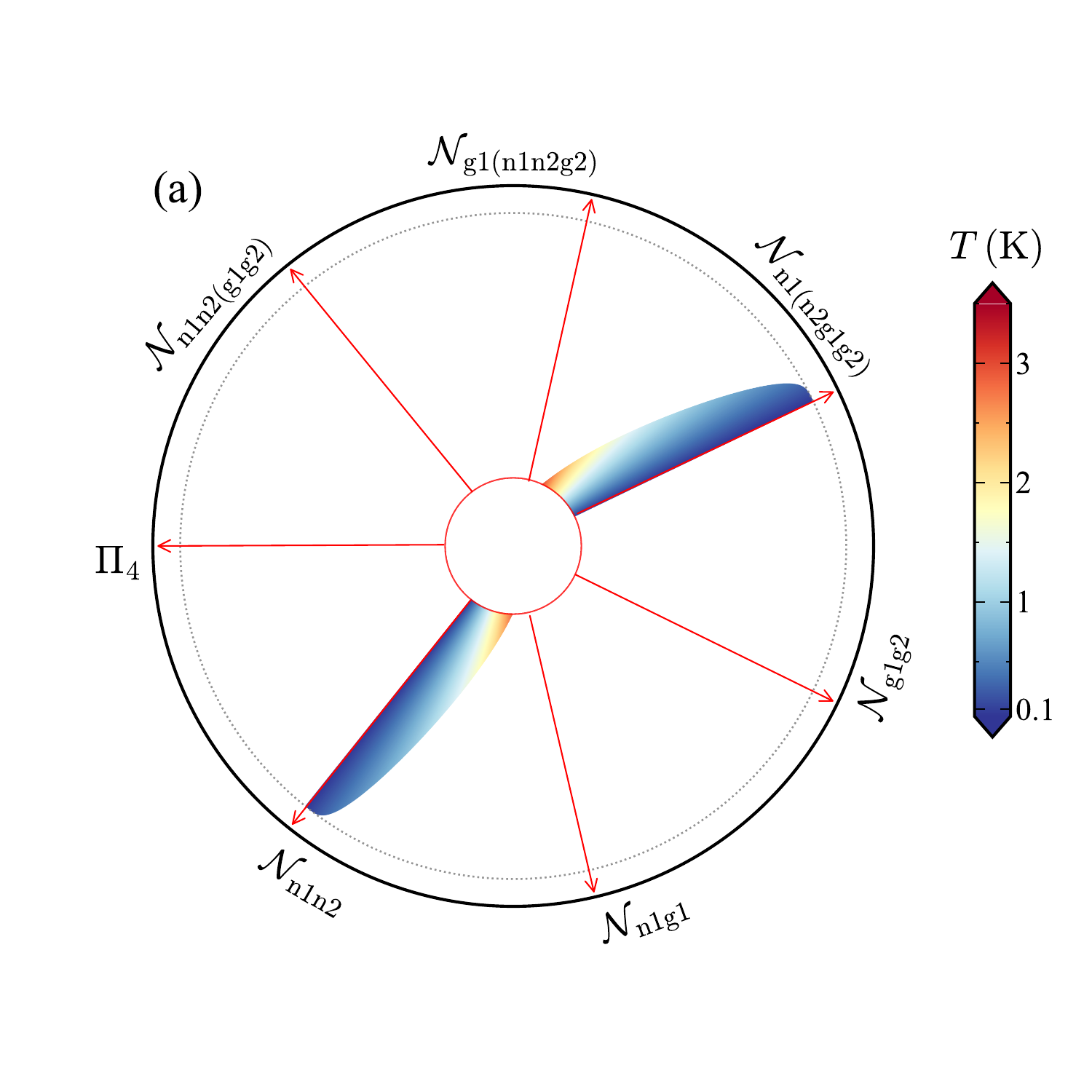}
	\includegraphics[scale=0.28,trim=20 100 20 80, clip]{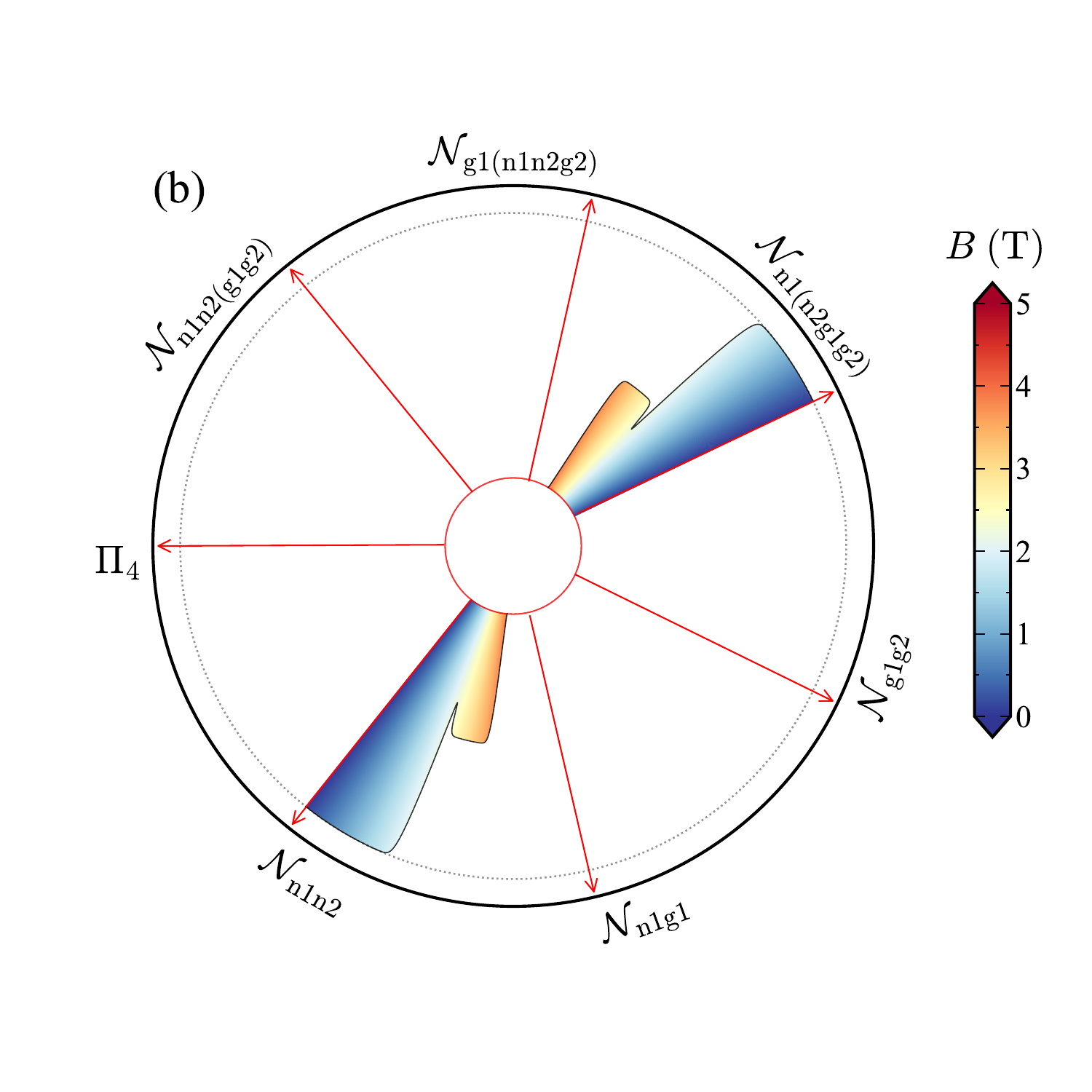}
	\includegraphics[scale=0.28,trim=20 100 10 60, clip]{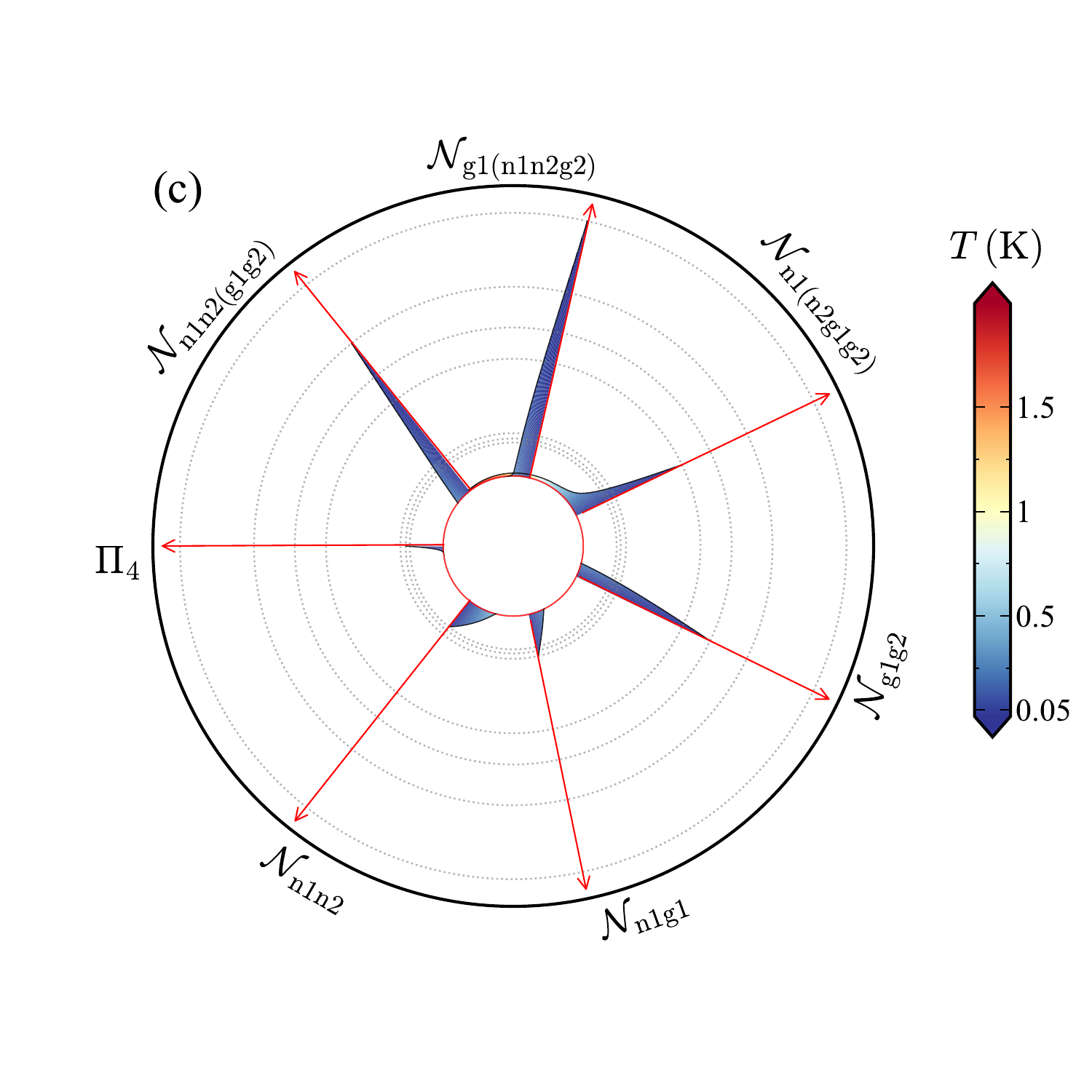}
	\includegraphics[scale=0.28,trim=20 100 20 60, clip]{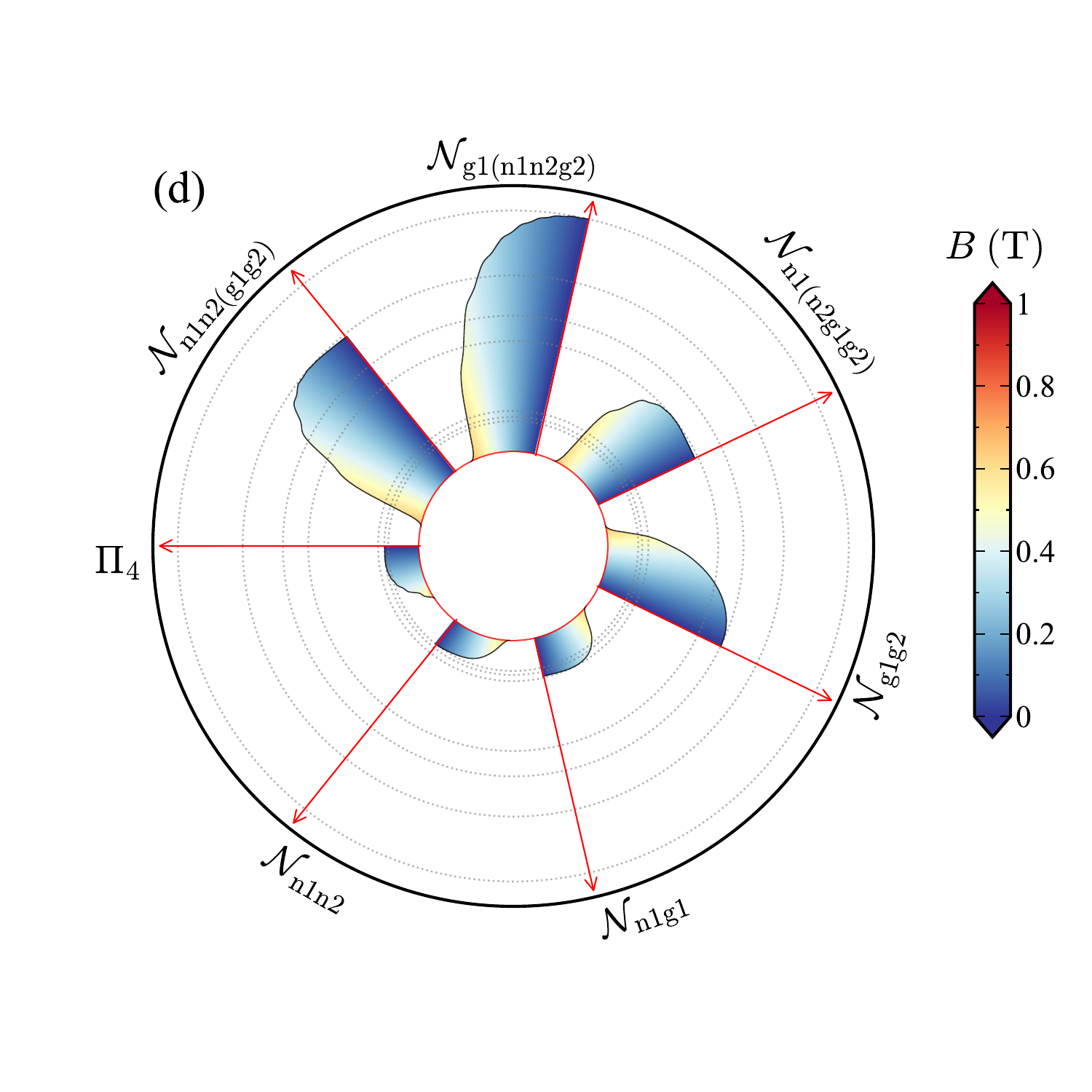}
	\caption{
		Bar charts illustrate various quantum correlations, including tetrapartite entanglement $\Pi_4$, bipartite negativities $\mathcal{N}_\mathrm{n1n2}$, $\mathcal{N}_\mathrm{n1g1}$, and $\mathcal{N}_\mathrm{g1g2}$, as well as the 1--3 tangles $\mathcal{N}_\text{n1(n2g1g2)}$ and $\mathcal{N}_\text{g1(n1n2g2)}$, and the 2--2 tangle $\mathcal{N}_\text{n1n2(g1g2)}$.
		(a) For complex \textbf{(1)} as a function of temperature at $B = 0$, and
		(b) for complex \textbf{(1)} as functions of magnetic field at $T=0.07\,\text{K}$. The Hamiltonian parameters are:
		$D_\mathrm{n} = 0$, $J_\mathrm{nn} = 1.53\;\mathrm{cm}^{-1}$, $J_\mathrm{ng} = 0.0074\;\mathrm{cm}^{-1}$, $J_\mathrm{gg} = -0.288\;\mathrm{cm}^{-1}$, $g_\mathrm{n} = g_\mathrm{g} = 2.0$, and inter-cubane interaction $zJ' = 0.0013\;\mathrm{cm}^{-1}$.
		(c) For complex \textbf{(2)} as functions of temperature at $B = 0$, and
		(d) for complex \textbf{(2)} as a function of magnetic field at $T=0.07\,\text{K}$, assuming the Hamiltonian parameters:
		$D_\mathrm{n} = 2.1$, $J_\mathrm{nn} = -5.2\;\mathrm{cm}^{-1}$, $J_\mathrm{ng} = -0.86\;\mathrm{cm}^{-1}$, $J_\mathrm{gg} = 0.0034\;\mathrm{cm}^{-1}$, $g_\mathrm{n} = g_\mathrm{g} = 2.1$, and $zJ' = -0.0002\;\mathrm{cm}^{-1}$.
		Dotted gray circles indicate the maximum value attained by each quantity as mentioned in Table \ref{tab:I}. Red radial arrows denote sector boundaries for each quantity (analogous to the $y$-axis), while the red central circle represents the baseline temperature or field vector. The actual values of the temperature  (field) axis are encoded by the color scale shown in the accompanying color bar.
	}
	\label{fig:barcharts}
\end{figure*}
 
\begin{figure}
	\includegraphics[scale=0.3,trim=10 0 20 50, clip]{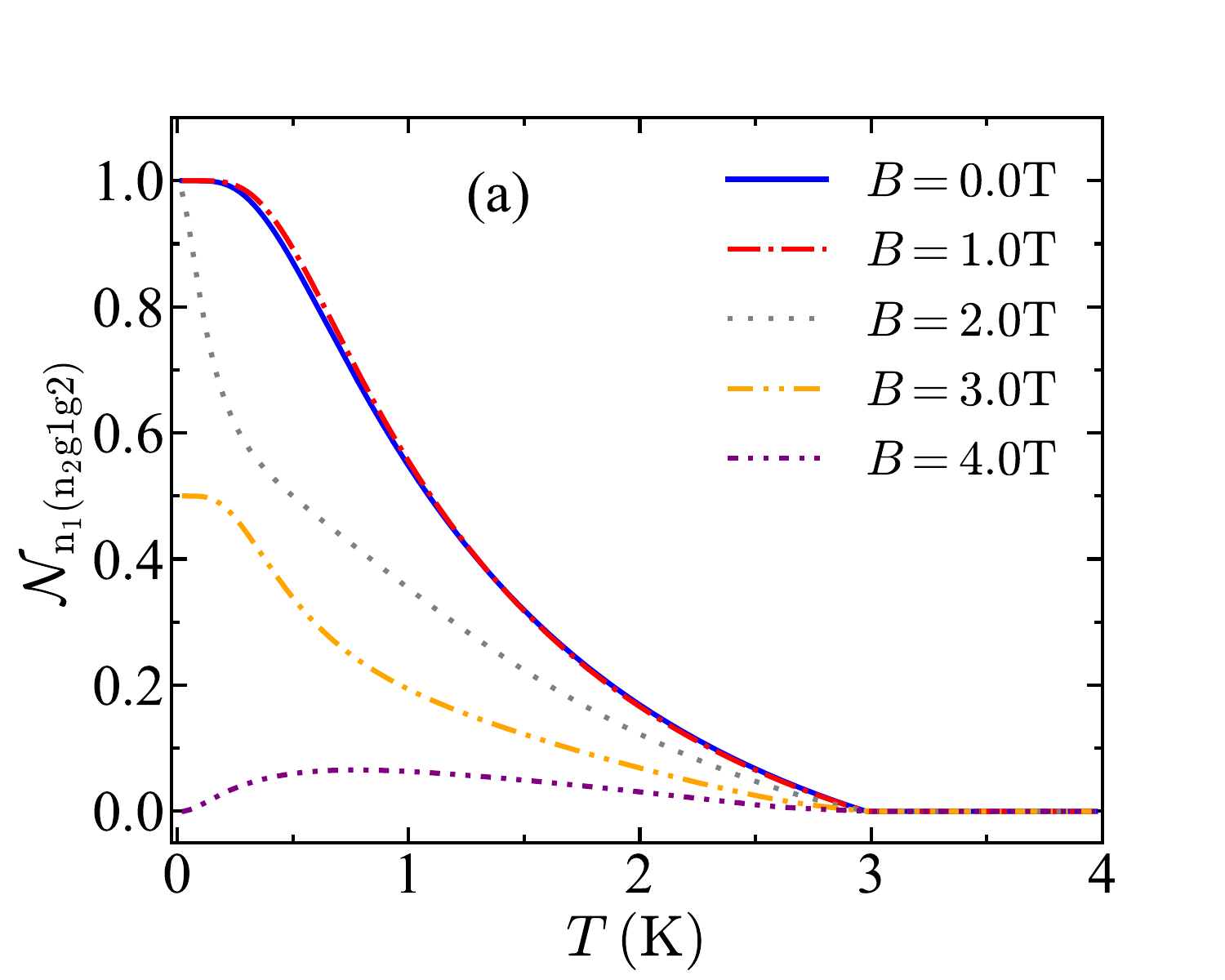}
	\includegraphics[scale=0.3,trim=0 0 20 50, clip]{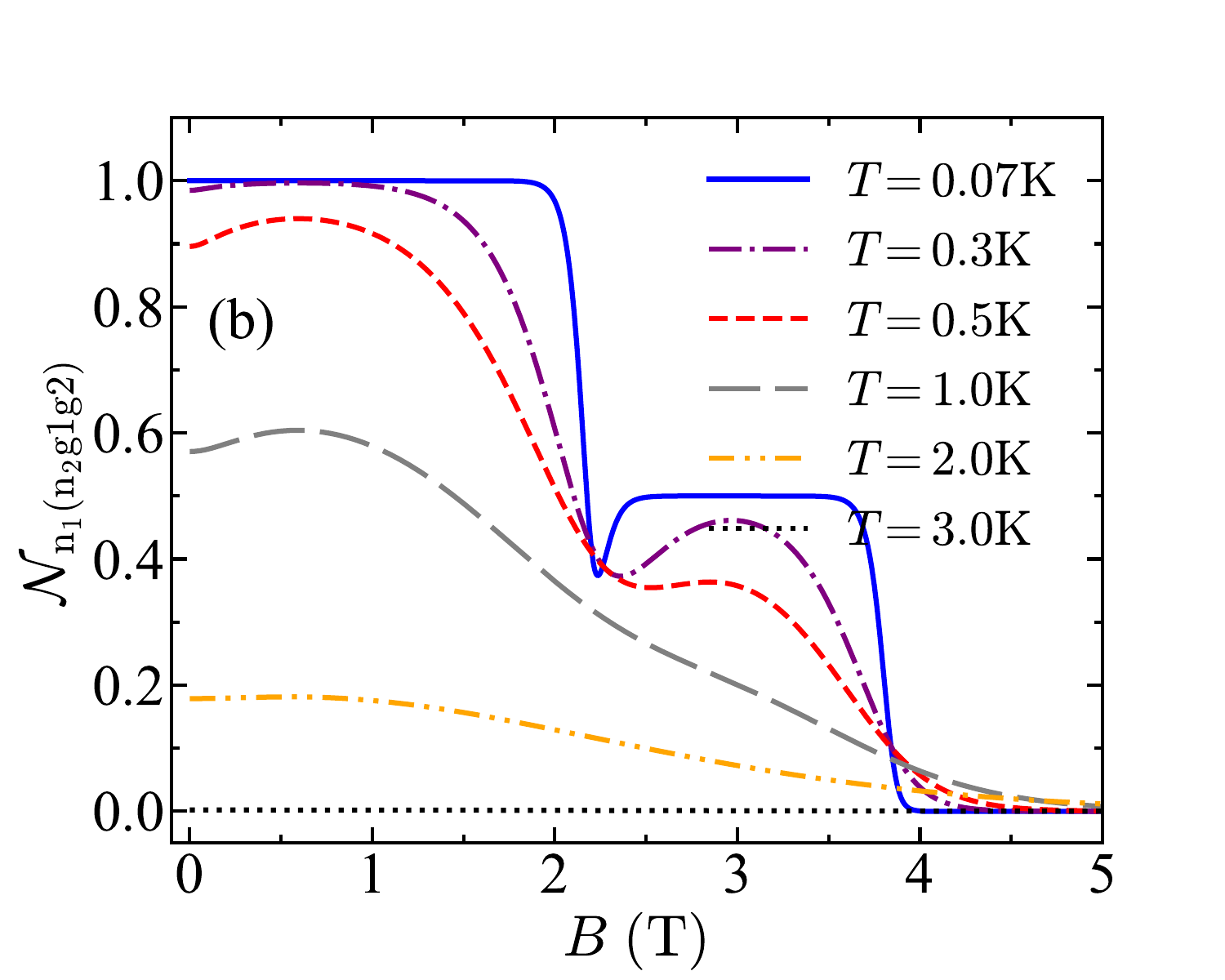} %
	\caption{  
		(a) Temperature dependence of the 1--3 tangle negativity, \(\mathcal{N}_\text{n1(n2g1g2)}\), for the complex {\bf (1)} described by Hamiltonian (\ref{eq:hamiltonian}) with parameters: \( D_\text{n} = 0 \), \( J_\text{nn} = 1.53\;\text{cm}^{-1} \), \( J_\text{ng} = 0.0074\;\text{cm}^{-1} \), \( J_\text{gg} = -0.288\;\text{cm}^{-1} \), $g_\text{n} = g_\text{g} = 2.0$ and  inter-cubane interaction \( zJ^{\prime} = 0.0013\;\text{cm}^{-1} \).  
		(b) Magnetic field dependence of \(\mathcal{N}_\text{n1(n2g1g2)}\) at various temperatures for the same parameter set as in panel (a). It is worth noting that, here, \(\mathcal{N}_\text{n1(n2g1g2)}\) behaves identically to \(\mathcal{N}_\text{n1n2}\).
	}
	\label{fig:bipneg_Biswas}
\end{figure}

\begin{table}[h]
	\centering
	\caption{Maximum values of various quantum entanglement negativities for complexes \textbf{(1)} and \textbf{(2)} at low temperature $T = 0.07\,\text{K}$. These values correspond to the radii of the dotted reference circles shown in Fig.~\ref{fig:barcharts}.}
	\begin{tabular}{l@{\hspace{5em}}c@{\hspace{5em}}c}
		\hline
		\hline
		\textbf{Negativity} & complex \textbf{(1)} & complex \textbf{(2)} \\
		\hline
		$\Pi_4$ & 0 & $0.077$ \\
		$\mathcal{N}_{\mathrm{n1n2}}$ & $1.0$ & $0.0687$ \\
		$\mathcal{N}_{\mathrm{n1g1}}$ & 0 & $0.090$ \\
		$\mathcal{N}_{\mathrm{g1g2}}$ & 0 & $0.295$ \\
		$\mathcal{N}_{\mathrm{n1(n2g1g2)}}$ & $1.0$ & $0.240$ \\
		$\mathcal{N}_{\mathrm{g1(n1n2g2)}}$ & 0 & $0.521$ \\
		$\mathcal{N}_{\mathrm{n1n2(g1g2)}}$ & 0 & $0.381$ \\
		\hline
		\hline
	\end{tabular}\label{tab:I}
\end{table}
Figure~\ref{fig:barcharts} presents bar charts of seven quantum entanglement measures, including the tetrapartite entanglement $\Pi_4$, bipartite negativities $\mathcal{N}_{\mathrm{n1n2}}$, $\mathcal{N}_{\mathrm{n1g1}}$, $\mathcal{N}_{\mathrm{g1g2}}$, $\mathcal{N}_{\mathrm{n1(n2g1g2)}}$, $\mathcal{N}_{\mathrm{g1(n1n2g2)}}$, and  $\mathcal{N}_{\mathrm{n1n2(g1g2)}}$, in a single cubane unit $\text{Ni}^{2+}_2\text{Gd}^{3+}_2$ of the two molecular complexes \textbf{(1)} and \textbf{(2)}.
Radius of the dotted circles marks the maximum value attained by each quantity at $T = 0.07\,\text{K}$, as reported in Table~\ref{tab:I}.
Red radial arrows indicate hight of the quantities on the $y$-axis, and the red central circle represents the vector for temperature or magnetic field on the $x$-axis. The actual values of the temperature or field axis are encoded in the accompanying color scale.
Figure~\ref{fig:barcharts}(a) shows the entanglement quantities for complex \textbf{(1)} as a function of temperature at zero magnetic field, using the Hamiltonian parameters from Fig.~\ref{fig:MagExp_ED}(c). As discussed in the previous section, for complex \textbf{(1)}, the possible ground states are VII and VIII described in Eq. (\ref{eq:VIVIIVIII_1}). In these states, the only significant correlations occur between Ni$\cdots$Ni ions. Consequently, only the bipartite negativity $\mathcal{N}_{\mathrm{n1n2}}$ and the 1--3 tangle $\mathcal{N}_{\mathrm{n1(n2g1g2)}}$ are equal and nonzero, while all other entanglement measures are vanished. In Fig. \ref{fig:bipneg_Biswas}(a), we depict the negativity $\mathcal{N}_{\mathrm{n1(n2g1g2)}}$ of \textbf{(1)} as a function of temperature for different values of the magnetic field. A slight increase in negativity is observed for magnetic field values in the range $0 < B \lesssim 1\,\text{T}$, followed by a marked decrease as the field strength increases further.

Figure~\ref{fig:barcharts}(b) displays the negativities for complex \textbf{(1)} as a function of magnetic field at low temperature $T = 0.07\,\mathrm{K}$. Again, only $\mathcal{N}_{\mathrm{n1n2}}$ and $\mathcal{N}_{\mathrm{n1(n2g1g2)}}$ are equally nonzero. As the magnetic field increases, both measures decrease from their maximum value of unity (corresponding to the state VII, as indicated by the dotted circle with radius $r = 1$), then reach a plateau at intermediate values corresponding to the VIII state, and finally vanish at high field strengths $B>4\,\text{T}$ where the system is fully polarized. More clear field dependence of the negativity $\mathcal{N}_{\mathrm{n1(n2g1g2)}}$ for various fixed temperatures is shown in Fig. \ref{fig:bipneg_Biswas}(b). A gradual decrease of $\mathcal{N}_{\mathrm{n1(n2g1g2)}}$ is observed when the temperature increases. 
At low temperature ($T<0.3\,\text{K}$), upon strengthening the field, this quantity starts to decrease from a wide plateau at $\mathcal{N}_{\mathrm{n1n2}}=1.0$ and drops down close to the transition field $B\approx 2.1\,\text{T}$ where the state of the system changes from VII to VIII, then slightly increases and passes a narrower plateau and finally shows another decline close to the second transition field $B\approx 3.6\,\text{T}$ at which the ground-state phase transition occurs between the state VIII with magnetization $M/M_\text{s} = \frac{8}{9}$ and fully polarized one IX (see Fig. \ref{fig:MagExp_ED}(a)). As we previously discussed in Ref. \cite{ArianPRA2025}, here a prominent suppression of the negativity occurs precisely at the coexistence point of the ground states VII and VIII. This effect originates from the formation of a mixed quantum state at the phase boundary. The mixedness of this state leads to a substantial reduction in the entanglement negativity below the value observed in the ground state VIII. 

Figures~\ref{fig:barcharts}(c) and (d) depict the negativities for complex \textbf{(2)} as functions of temperature at zero magnetic field and as a function of magnetic field at $T = 0.07\,\mathrm{K}$, respectively, using the Hamiltonian parameters from Fig.~\ref{fig:MagExp_ED}(d). A clear contrast is observed between complexes \textbf{(1)} (with zero anisotropy) and \textbf{(2)} (with nonzero anisotropy): in complex \textbf{(2)}, none of the entanglement measures vanish. 

\begin{figure*}
	\includegraphics[scale=0.22,trim=10 0 50 50, clip]{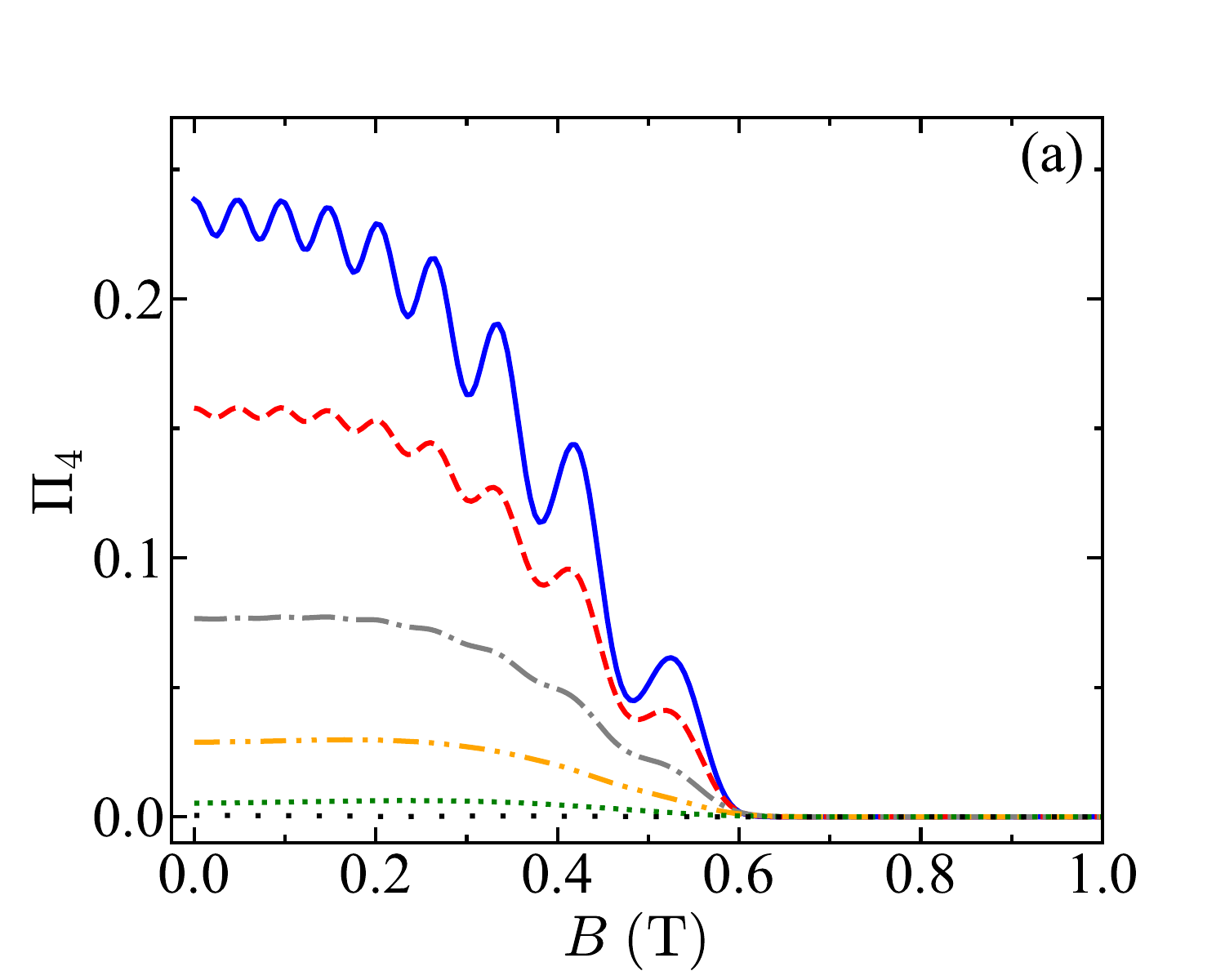}
	\includegraphics[scale=0.22,trim=10 0 50 50, clip]{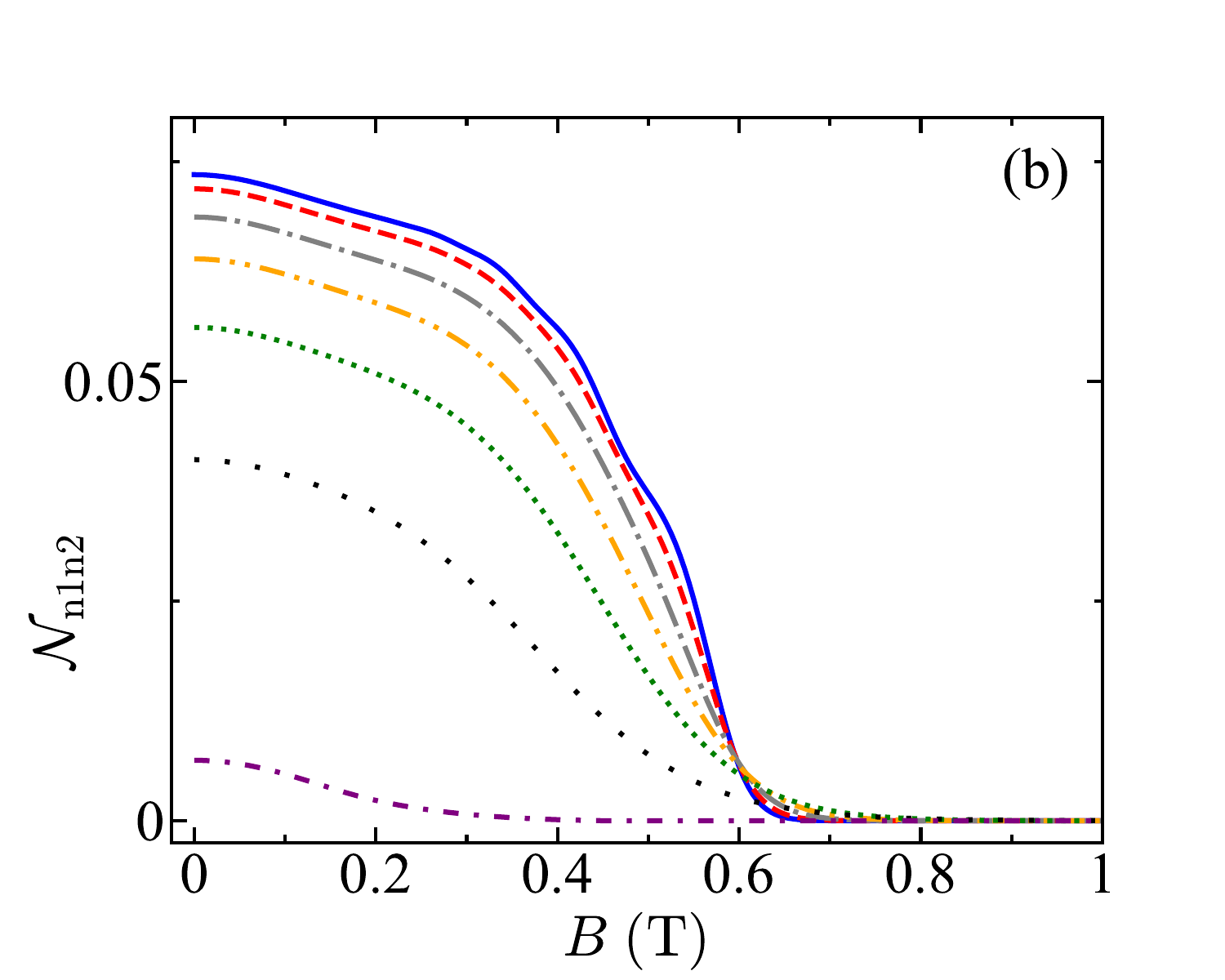} %
	\includegraphics[scale=0.22,trim=10 0 50 50, clip]{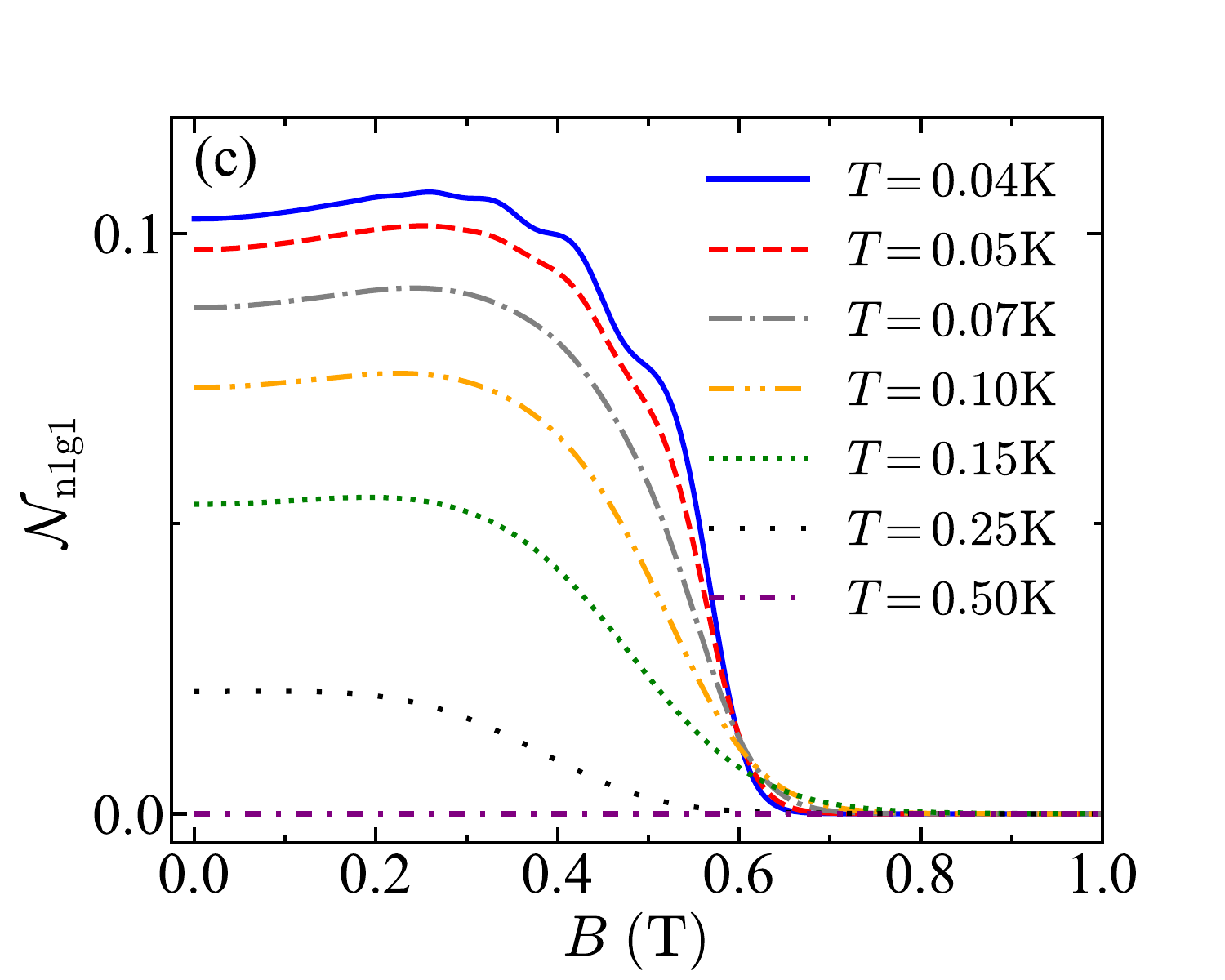}
	\includegraphics[scale=0.22,trim=10 0 50 50, clip]{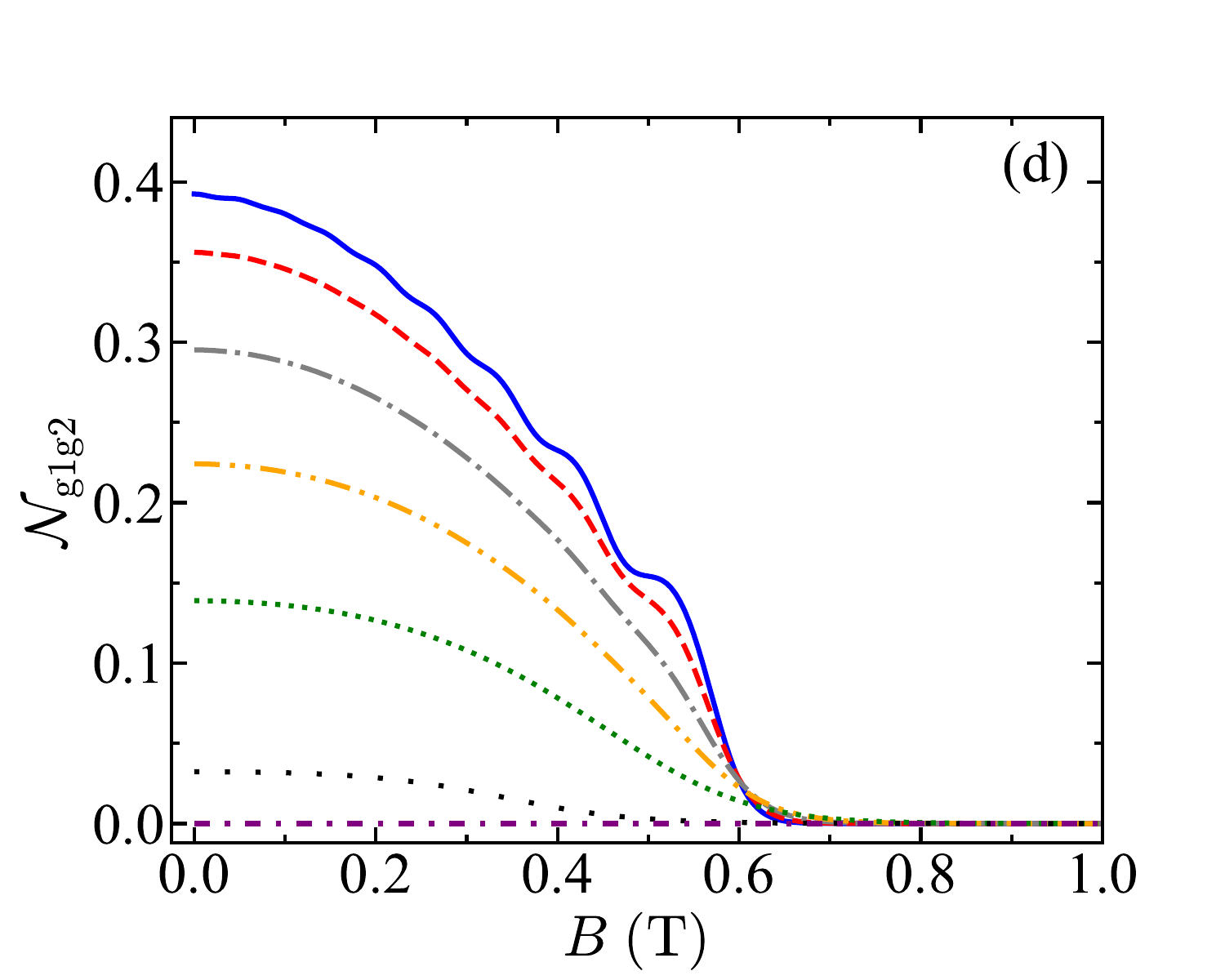}
	\includegraphics[scale=0.22,trim=10 0 50 50, clip]{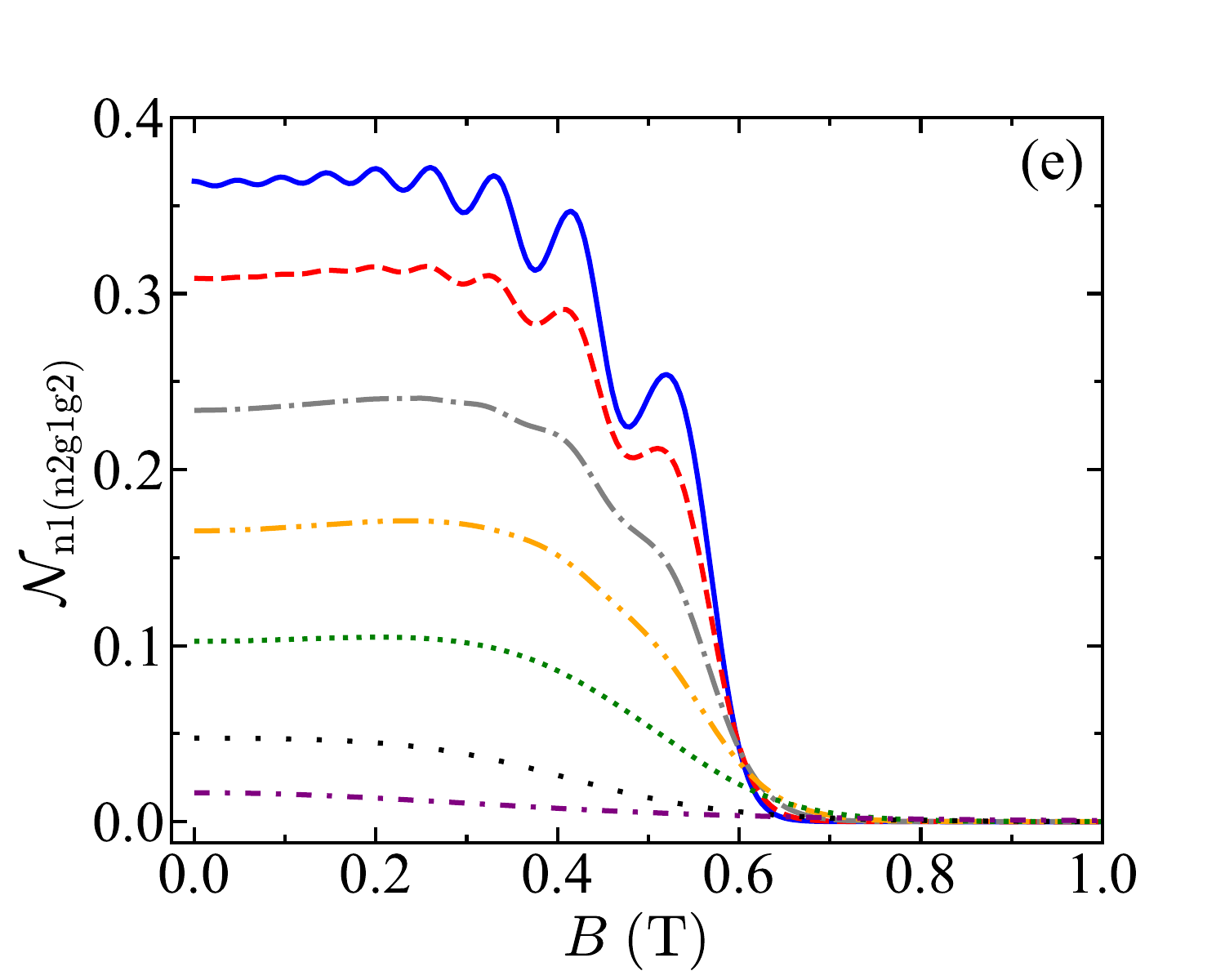}
	\includegraphics[scale=0.22,trim=10 0 50 50, clip]{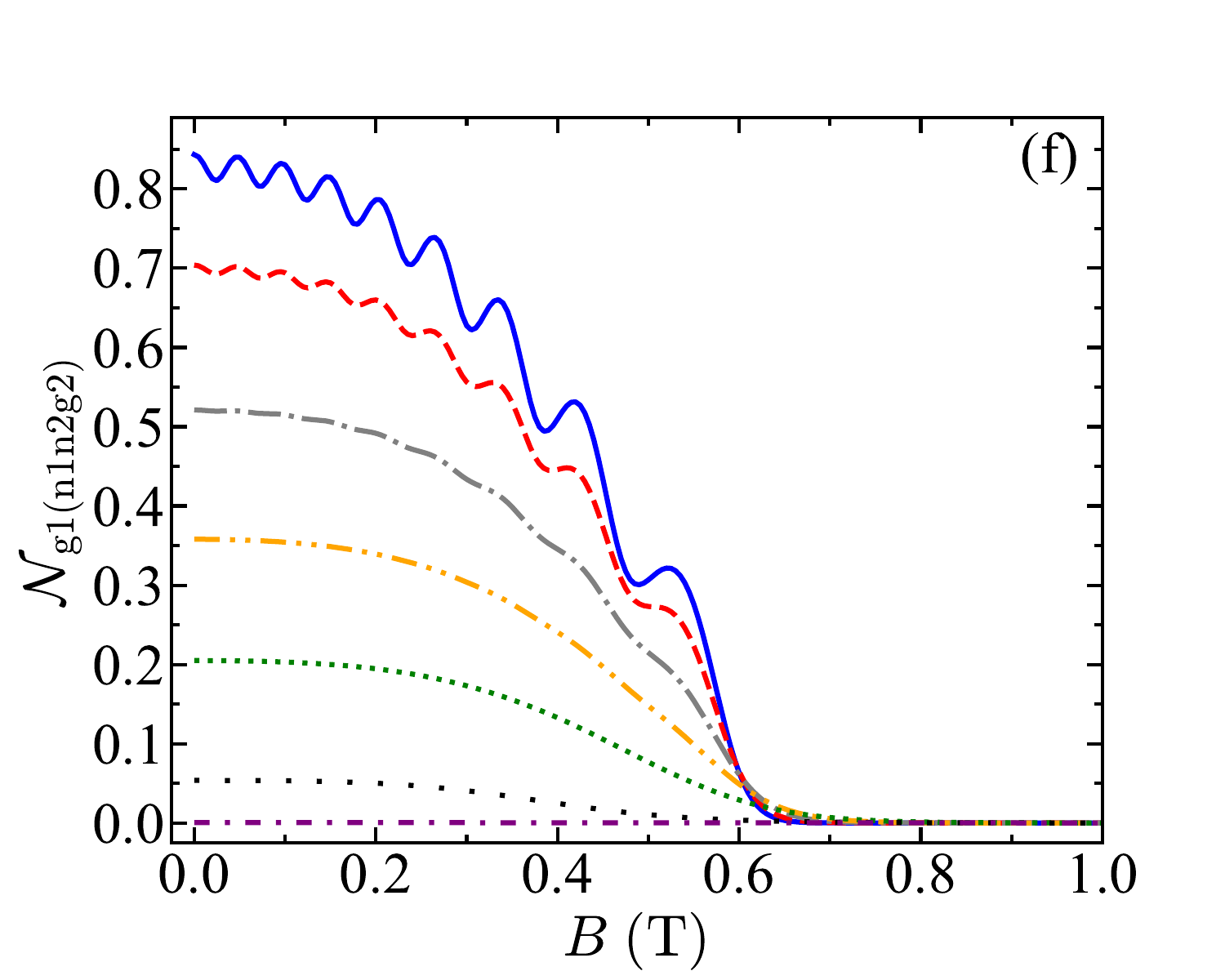}
	\includegraphics[scale=0.22,trim=10 0 50 50, clip]{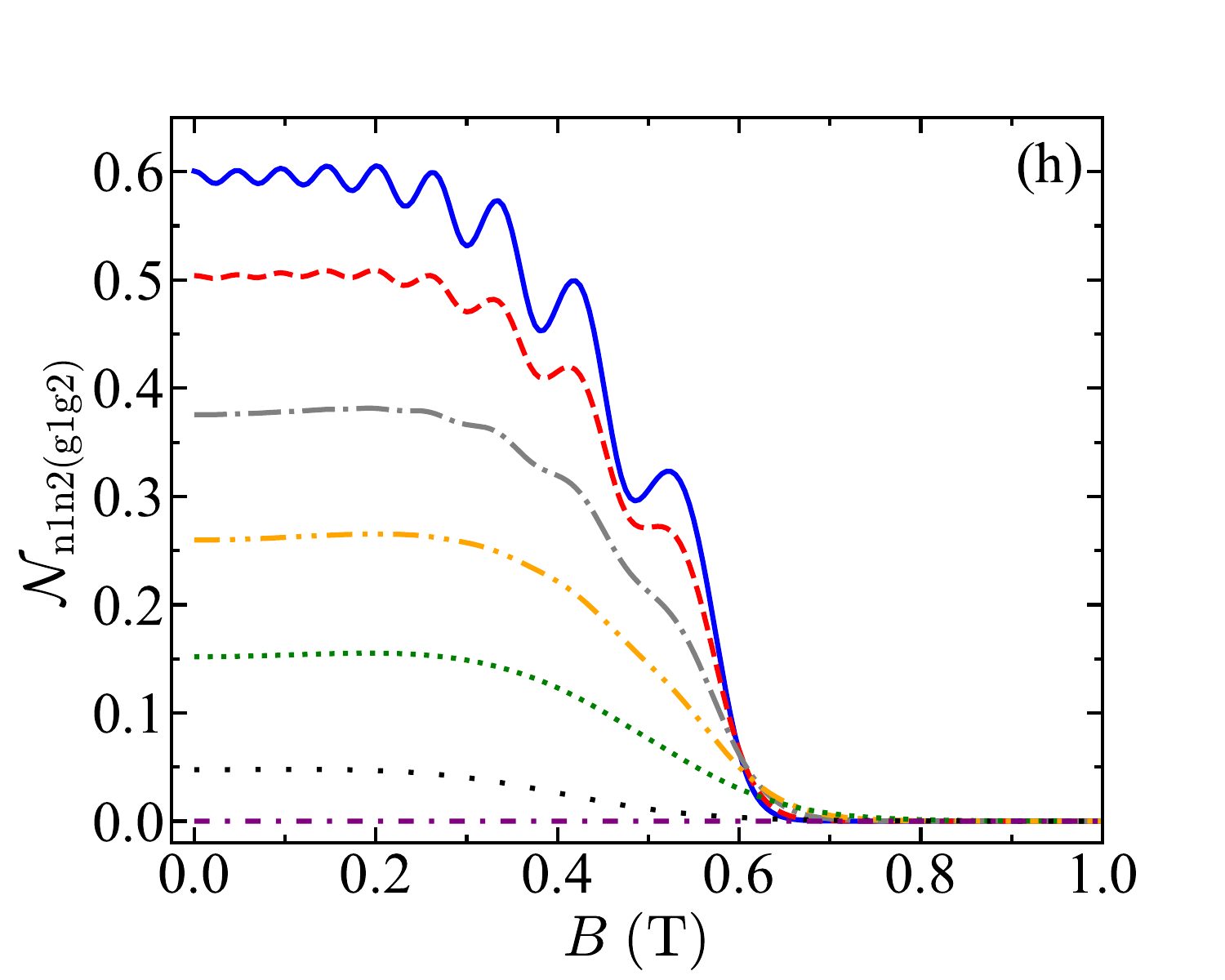}
	\caption{  
		 Field dependence of the entanglement negativities for the complex {\bf (2)} at various fixed temperatures, considering the same parameters as in panel Fig. \ref{fig:MagExp_ED}(d).
		(a) The whole entanglement $\Pi_4$,
		(b) $\mathcal{N}_\text{n1n2}$, 
		(c) $\mathcal{N}_\text{n1g1}$,  
		(d) $\mathcal{N}_\text{g1g2}$,
		(e) $\mathcal{N}_{\mathrm{n1(n2g1g2)}}$, 
		(f) $\mathcal{N}_{\mathrm{g1(n1n2g2)}}$, and 
		(h) 2--2 tangle $\mathcal{N}_{\mathrm{n1n2(g1g2)}}$.
		}
	\label{fig:AllNegs_Kalita}
\end{figure*}

The thermal quantum entanglement negativities of a cubane unit of the complex {\bf (2)} are illustrated in Fig. \ref{fig:AllNegs_Kalita}.
It is quite evident from Fig. \ref{fig:AllNegs_Kalita}(a) that for complex \textbf{(2)}, the whole entanglement $\Pi_4$ is nonzero and is relatively strong. The main reason is due to nonzero single-ion anisotropy $D_\text{n}$ of the Ni ions.
To assess whether the entanglement negativities depicted in Fig.~\ref{fig:barcharts}(d) effectively capture the field-induced phase transition of complex~\textbf{(2)}, we examine these quantities at lower temperatures than $T = 0.07\,\mathrm{K}$. Specifically, Fig.~\ref{fig:AllNegs_Kalita} includes the entanglement negativities as functions of the magnetic field at $T = 0.04\,\mathrm{K}$ (blue solid line) and $T = 0.05\,\mathrm{K}$ (red dashed line).
It is evident that, entanglement measures including  $\Pi_4$, the 1--3 tangles $\mathcal{N}_{\mathrm{n1(n2g1g2)}}$ and $\mathcal{N}_{\mathrm{g1(n1n2g2)}}$, and the 2--2 tangle $\mathcal{N}_{\mathrm{n1n2(g1g2)}}$ plausibly reflect the  ground-state phase transition shown in Fig.~\ref{fig:MagExp_ED}(b), as well as the zero-temperature magnetization behavior illustrated in Fig.~\ref{fig:MagExp_ED}(d).
Despite the strong ferromagnetic Ni$\cdots$Ni interaction and the relatively weaker ferromagnetic Ni$\cdots$Gd coupling, a small yet non-negligible degree of tetrapartite entanglement survives within the single cubane unit of complex~\textbf{(2)} for temperatures $T < 0.25\,\mathrm{K}$ and magnetic fields $B \lesssim 0.6\,\mathrm{T}$.


In Figs.~\ref{fig:AllNegs_Kalita}(b)--(d), we plot the bipartite negativities $\mathcal{N}_\text{n1n2}$, $\mathcal{N}_\text{n1g1}$, and $\mathcal{N}_\text{g1g2}$,  defined in Eqs.~(\ref{negativities_a})--(\ref{negativities_c}), as functions of the magnetic field for several fixed temperatures. These figures reveal that nonzero values of the bipartite entanglement negativities arise mainly due to the presence of single-ion anisotropy $D_\text{n}$. At low temperatures, these quantities generally exhibit a smooth decline with increasing magnetic field. 
In particular, the bipartite entanglement negativity $\mathcal{N}_\text{n1n2}$ shown in Fig.~\ref{fig:AllNegs_Kalita}(b) is significantly weaker than that observed in complex~(\textbf{1}) (see Fig.~\ref{fig:bipneg_Biswas}), which can be attributed to the strong ferromagnetic interaction between the Ni ions.
In contrast, the 1--3 tangles $\mathcal{N}_{\mathrm{n1(n2g1g2)}}$, $\mathcal{N}_{\mathrm{g1(n1n2g2)}}$, and the 2--2 tangle $\mathcal{N}_{\mathrm{n1n2(g1g2)}}$, shown in Figs.~\ref{fig:AllNegs_Kalita}(e)--(f), display a non-monotonic increase--decrease pattern at low temperatures $T \lesssim 0.05\,\mathrm{K}$, reminiscent of the ground-state phase transition depicted in Fig.~\ref{fig:MagExp_ED}(b). 

By comparing all the negativities, from Fig.~\ref{fig:AllNegs_Kalita}(f) one finds that the strongest entanglement occurs for the tangle $\mathcal{N}_{\mathrm{g1(n1n2g2)}}$, namely, between the Gd ion and the rest of the parties in the cubane.
Given that the exchange interactions between Ni$\cdots$Ni and Ni$\cdots$Gd in complex~(\textbf{2}) are ferromagnetic, one might intuitively expect a suppression of pairwise quantum correlations. However, the pronounced bipartite entanglement negativities observed indicate the involvement of an alternative mechanism. We attribute this enhancement to the presence of nonzero single-ion anisotropy in the Ni ions, which introduces quantum fluctuations that counteract the spin polarization imposed by the ferromagnetic exchange interactions.
 
{ Beyond the computational results, it is important to emphasize the experimental relevance of the entanglement negativities calculated here. Although negativity itself is not a direct observable, it has been demonstrated in recent works \cite{Str2020, vargova2021, Cramer2010, Cramer2011} that entanglement measures can be inferred from quantities accessible to experiment. Specifically, the elements of the reduced density matrix, on which negativity is based, are strongly correlated with local observables such as magnetization \( \langle \hat{S}_i^z \rangle \), pairwise spin-spin correlators \( \langle \hat{S}_i^z \hat{S}_j^z \rangle \), and two-spin expectation values \( \langle \vec{S}_i \cdot \vec{S}_j \rangle \), all of which can be extracted through inelastic neutron scattering, muon spin rotation (\(\mu\)SR), or electron spin resonance (ESR) measurements. Moreover, thermodynamic quantities like specific heat and magnetic susceptibility, which have already been used to validate the magnetic models in this study, implicitly encode correlations that reflect underlying entanglement properties. Thus, the field and temperature dependence of bipartite and multipartite negativities presented in this work—particularly their sharp changes across quantum phase boundaries—can serve as indirect indicators of critical behavior and entanglement dynamics in molecular magnets. These insights point toward a promising pathway for experimentally probing and controlling entanglement in transition-metal/lanthanide complexes, with potential implications for spin-based quantum information devices.}

These findings highlight the crucial role of the single-ion anisotropy in generating quantum entanglement even in systems where ferromagnetic interactions prevail over antiferromagnetic ones. This insight reveals the potential of non-SMMs as platforms for studying entanglement generation mechanisms beyond conventional antiferromagnetic coupling schemes. Furthermore, our results suggest that enhancing single-ion anisotropy could provide a viable strategy for engineering robust entanglement in molecular magnetic systems in particular heterometallic \(3d/4f\) complexes which paves the way for advancements in quantum information processing and quantum molecular magnetism.

\section{Conclusions}\label{conclusions}

In this work, we have rigorously investigated the ground-state phase transitions and low-temperature magnetic properties, { spin level structure}, along with the different types of entanglement negativities, of the { \(\text{Ni}_4^{2+}\text{Gd}_4^{3+}\)} molecular complexes under an external magnetic field. Our findings reveal that these complexes exhibit multiple ground states, each accompanied by commensurate magnetization plateaus.
Furthermore, our analysis of the quantum properties of these complexes indicates that while the tetrapartite entanglement in a single cubane of complex {\bf (1)} vanishes while in complex {\bf (2)} it is nonzero, the bipartite entanglement negativity $\mathcal{N}_\text{n1n2}$ displays considerably higher values. This apparent discrepancy emphasis the crucial role of single-ion anisotropy in generating and enhancing different types of entanglement within heterometallic \(3d/4f\) complexes including Ni ions, despite the dominance of ferromagnetic interactions.

The particular negativity $\mathcal{N}_{\mathrm{n1(n2g1g2)}}$ of these complexes exhibit anomalous behavior near the critical transition fields, reflecting the intricate quantum correlations that emerge in response to varying external magnetic field. This entanglement measure can be served as a reliable indicator of the ground-state phase transitions, as its behavior changes markedly with the strength of the magnetic field and single-ion anisotropy specifically across phase boundaries. The pronounced variation of this negativity near transition points highlight the interplay between quantum entanglement and magnetic ordering that provide a deeper insights into the fundamental mechanisms of the phase transition in such magnetic materials. 
We have also found that the tetrapartite entanglement negativity together with 1--3 tangles can be considered as applicable  indicators of phase transitions in { \(\text{Ni}_4^{2+}\text{Gd}_4^{3+}\)} complexes when the single-ion anisotropy is nonzero. 

Although the dominant ferromagnetic interactions between Ni$\cdots$Ni and Ni$\cdots$Gd in complex \textbf{(2)} might suggest a suppression of quantum correlations, our results indicate that the presence of nonzero single-ion anisotropy counteracts this effect by introducing quantum fluctuations, and consequently, give rise to various forms of entanglement. This highlights an important mechanism by which quantum entanglement can be engineered even in systems where ferromagnetic interactions dominate over antiferromagnetic ones.  

Our results have broader implications for the design and control of quantum entanglement in molecular magnets, particularly in heterometallic \(3d/4f\) complexes. The ability of single-ion anisotropy to generate and enhance robust entanglements provides a promising route for exploring novel quantum features in these magnetic compounds. Future studies may discover how varying the anisotropy or introducing additional external control parameters can further optimize entanglement in the similar families of molecular systems.   

{ Drawn from the specific case of \(\text{Ni}_4^{2+}\text{Gd}_4^{3+}\), our insights are not limited to this system. Rather, the core mechanism we identify serves as a quantitative roadmap: the strategic use of single-ion anisotropy to counteract the suppression of quantum correlations by ferromagnetic interactions. This underlying approach of leveraging competing interactions presents a general and powerful strategy for the design of entanglement in a broad class of isostructural \(3d/4f\) molecular magnets (e.g., Cu--Ln, Cr--Ln, Mn--Ln, and Co--Ln), with direct applications in quantum magnetism, quantum information science, and the broader pursuit of correlated phenomena in condensed matter physics.}

\section*{Acknowledgments}
H.A.Z. acknowledges the financial support provided under
the postdoctoral fellowship program of P. J. Šafárik University in Košice, Slovakia. This research was funded by the
Slovak Research and Development Agency under Contract
No. APVV-20-0150, and under the contract No. VVGS-2023-2888, and the Ministry of Education, Research,
Development and Youth of the Slovak Republic under Grant
No. VEGA 1/0298/25. N.A. acknowledge the receipt of the grant in the frame of the research projects No. SCS 21AG-1C006.


\end{document}